\documentclass[a4paper,fleqn,usenatbib]{mnras}

\usepackage{newtxtext,newtxmath}

\usepackage[T1]{fontenc}
\usepackage{ae,aecompl}

\usepackage{graphicx}	
\usepackage{amsmath}	
\usepackage{amssymb}	

\title[New full evolutionary sequences of massive WDs]{New full evolutionary sequences of H and He atmosphere massive white dwarf stars using MESA}

\author[G.R. Lauffer et al.]{
G.R. Lauffer,$^{1}$\thanks{E-mail: gabriel.lauffer@ufrgs.br}
A. D. Romero,$^{1}$
S.O. Kepler$^{1}$
\\
$^{1}$Instituto de F\'{i}sica, Universidade Federal do Rio Grande do Sul, Av. Bento Goncalves 9500, Porto Alegre 91501-970, RS, Brazil
}

\date{Accepted XXX. Received YYY; in original form ZZZ}

\pubyear{2018}

\begin{document}
\label{firstpage}
\pagerange{\pageref{firstpage}--\pageref{lastpage}}
\maketitle

\begin{abstract}
We explore the evolution of hydrogen-rich and hydrogen-deficient white dwarf stars with masses between $1.012$ and $1.307$ M$_{\rm \odot}$, and initial metallicity of $Z=0.02$. These sequences are the result of main sequence stars with masses between $8.8$ and $11.8$ M$_{\rm \odot}$.
The simulations were performed with MESA, starting at the zero-age main sequence, through thermally pulsing and mass-loss phases, ending at the white dwarfs cooling sequence.
We present reliable chemical profiles for the whole mass range considered, covering the different expected central compositions, i.e. C/O, O/Ne and Ne/O/Mg, and its dependence with the stellar mass. In addition, we present detailed chemical profiles of hybrid C/O-O/Ne core white dwarfs, found in the mass range between $1.024$ and $1.15$ M$_{\rm \odot}$.
We present the initial-to-final mass relation, mass-radius relation, and cooling times considering the effects of atmosphere and core composition.
\end{abstract}




\section{Introduction}

White dwarf stars are the most common final stage of stellar evolution. They are the end product of main sequence stars with masses up to $5-12$ M$_{\rm \odot}$ \citep{Garcia-berro97-2,Poelarends2008,Siess2010,Langer2012,Woosley2015,Doherty2015}, depending on metallicity, which correspond to $95-97\%$ of all stars in the Galaxy.
White dwarf stars are excellent galactic chronometers as they are slowly cooling, releasing the heat stored in the core. The age of stellar populations can be obtained by comparisons of theoretical computations with their distribution in the color-magnitude diagram \citep{Hansen2007,Campos2016} or from the white dwarf luminosity function \citep{Winget1987,Bedin2010,Garcia-berro2014}.
The connection between the properties of progenitor stars and white dwarfs is printed in the initial-to-final mass relation (IFMR) which leads to constraints on the upper mass limit that separates white dwarfs progenitors of Type II supernovae \citep{Siess2007,Catalan2008}. Also binary systems with white dwarfs are believed to produce Type Ia supernovae as result of accretion that exceeds the upper mass limit (e.g. \citealt{Wang2017}). Finally, a comparison between models and observations of white dwarfs can enhance the comprehension of the physical properties of high density matter (e.g. \citealt{PradaMoroni2002,Isern2010}).

The white dwarf stars can be divided into two main classes: DA and non-DA types, based on the main component of their atmosphere. The DA spectral class shows an hydrogen rich atmosphere and represent $\simeq 84 \%$ of all white dwarfs (e.g. \citealt{Kepler2007}) while non-DA's have hydrogen deficient atmospheres. The non-DA white dwarfs can be classified by the dominant element in the spectra and effective temperature: DO class with strong lines of HeII and effective temperatures of $T_{\rm eff} \approx 45,000 - 200,000$ K, DB class with strong HeI lines and $T_{\rm eff} \approx 11,000 - 30,000$ K, and for effective temperatures bellow $T_{\rm eff} \approx 11,000$ K there are DC, DQ and DZ types with traces of carbon and metals in their spectra.
The helium-rich white dwarfs are believed to be progeny of PG 1159 stars \citep{McGraw1979} formed by a Born-Again episode \citep{Schoenberner1979,iben1983} in which a helium flash occurs during the cooling sequence (i.e. very late thermal pulse) forcing the star to evolve back to the asymptotic giant branch (AGB). As a result, this very late thermal pulse burns out all hydrogen left in the atmosphere of the star.

White dwarfs with masses in the range $0.4 - 1.05$ M$_{\rm \odot}$ are believed to harbor a carbon-oxygen (C/O) core, after central hydrogen and helium burning on earlier stages of evolution. On the other hand, the progenitors of white dwarfs with masses larger than $\approx 1.05$ M$_{\rm \odot}$ \citep{Garcia-berro97-2,Siess2007} should reach temperature high enough to achieve stable carbon burning, forming an oxygen-neon (O/Ne) or neon-oxygen-magnesium (Ne/O/Mg) core, depending if carbon burning ignites off-center or at the center of the star. When carbon burning starts off-center, ignition is followed by a inward propagation of the burning front (hereafter carbon flame). This carbon flame may or may not reach the center. If the flame does not reach the center a hybrid C/O+O/Ne white dwarf will form \citep{Denissenkov2013,Farmer2015}, whilst a flame which reaches the center will form a Ne/O/Mg white dwarf. These results are dependent on the evolutionary code and input physics considered. \citet{Siess2007} calculated sequences for several metallicities, with and without overshooting, and found that for $Z=0.02$ the minimum core mass for O/Ne/ cores is $1.04$ M$_{\rm \odot}$.  \citet{Doherty2015}, based on sequences calculated until the thermally pulsing asymptotic giant branch phase (TP-AGB), found that for $Z=0.02$ the minimum core mass for a O/Ne core white dwarf star is $1.154$ M$_{\rm \odot}$.
The determination of the mass range for each type of core composition is fundamental to improve our understanding of the populations of massive white dwarf stars.

The number of spectroscopically confirmed white dwarf stars has increase to more than $\approx 30\, 000$ with surveys like the Sloan Digital Sky survey \citep{Anh2014}. The mass distribution for DA white dwarf stars presented by \citet{Kleinman2013} is centered at $0.649$ M$_{\rm \odot}$ but presents another peak at $0.811$ M$_{\rm \odot}$ \citep{Kepler2015}. In addition, an analysis of $112$ isolated massive white dwarfs performed by \citet{Nalezyty2004} showed that, for known white dwarfs with masses $\geq 0.8$ M$_{\rm \odot}$, the mass distribution has a primary peak at $\approx 0.8$ M$_{\rm \odot}$ and a secondary peak around $1.04$ M$_{\rm \odot}$.
Around $20\%$ of all DA white dwarfs are more massive than $0.8$ M$_{\rm \odot}$ \citep{Liebert2005,Kepler2007}
and about $9\%$ have mass around $1.12$ M$_{\rm \odot}$ \citep{Kepler2007}.
Recent works presents a high mass tail on the mass distribution with an excess for masses around $M_{\rm WD} \approx 1.0$ M$_{\rm \odot}$ \citep{Falcon2010,Tremblay2013,Rebassa-Mansergas2015,Kepler2016}.

In the literature, several evolutionary models for low and intermediate mass white dwarf stars can be found. For instance \citet{Romero2012,Romero2013} and \citet{Althaus2009} for C/O core DA and non-DA white dwarfs, respectively, and \citet{althaus2013,Istrate2014,Istrate2016,Istrate2017,Sun2017} for low and extremely low mass white dwarfs, are just some recent works.
On the other hand, there is a lack of full evolutionary models for massive white dwarfs. Some recent work computed the evolution of super asymptotic branch stars until the end of the thermally pulsing phase \citep{Siess2010,GilPons2013,Jones2013,Doherty2015}, but they do not compute the following post-AGB stage nor the white dwarf evolution. Other authors computed white dwarf models by generating static/politropic models at the top of the cooling sequence and/or by rescaling the mass and/or artificially inserting a previously calculated chemical profile with fixed O/Ne ratios and fixed hydrogen and helium layers on the atmosphere \citep{Benvenuto1999,Salaris2000,Althaus2005,Althaus2007,Salaris2013}.

The present work is intended to fill the gap due to the lack of full evolutionary sequences for massive white dwarfs, computed from the Zero Age Main Sequence (ZAMS) through the red giant branch (RGB) and AGB to the white dwarf cooling sequence. To this end we used the MESA evolutionary code \citep{Paxton2011,Paxton2013,Paxton2015,Paxton2017}. We computed full evolutionary sequences with initial masses in the range of $8.8 - 11.8$ $M_{\rm \odot}$ resulting in white dwarfs models with masses in the range of $1.01 - 1.31$ M$_{\rm \odot}$, adding both hydrogen and helium-rich envelopes.
This paper is organized as follows. On section \ref{sec:2-numericalsimulations} we present the MESA evolutionary code and the input physics considered in our computations. The results from our full evolutionary computations are presented in section \ref{sec:3-mass_evolution_before_cs}. Section \ref{sec:mass_radius} discusses the mass-radius relation of our sequences. Finally we present our final remarks on section \ref{sec:7-conclusion}.

\section{Numerical Simulations}
\label{sec:2-numericalsimulations}

The numerical simulations where performed using the MESA code \citep{Paxton2011, Paxton2013, Paxton2015,Paxton2017} version r8845. We started the computations on the ZAMS, considering an initial metallicity $Z=0.02$. We computed the hydrogen burning and helium burning stages, and the RGB and AGB mass loss phases. The sequences ended as C/O or O/Ne/Mg white dwarfs cooling until $\log L/L_{\rm \odot} \sim -4$, which represents $T_{\rm eff} \approx 7\, 000 - 10\, 000$ K depending on final mass.

The MESA evolutionary code has been extensively used to perform calculations of extremely low-mass white dwarfs \citep{Istrate2016,Istrate2017,Sun2017}, hybrid C/O/Ne white dwarfs and Type Ia SN progenitors \citep{Jones2013,Denissenkov2013,Chen2014,Farmer2015,Brooks2017}, accreting white dwarf binaries with C/O core \citep{Brooks2016,Wang2017} and O/Ne core \citep{josiah2017,Brooks2017-2}, white dwarf isochrones \citep{Dotter2016} and sets of sequences covering the white dwarf mass range \citep{Choi2016,Nugrid2016,Nugrid2017}.

We detail the input physics considered prior and during white dwarf evolution in the following subsections.

\subsection{Pre-white dwarf evolution}

On the evolutionary phases prior to the cooling sequence, we considered the nuclear network \texttt{co\_burn\_plus.net} which has 16 elements from hydrogen to silicon, $^1$H, $^{3,4}$He, $^{12,13}$C, $^{13-15}$N, $^{16-18}$O, $^{19}$F, $^{20, 22}$Ne, $^{24}$Mg, and $^{28}$Si.
This network has 67 nuclear reactions, including pp-chain, CNO cycle, triple-alpha and some auxiliary reactions.
We used JINA reaclib \citep{Cyburt2010} with electron screening factors for thermonuclear reactions from \citet{Graboske1973} and \citet{DeWitt1973} for weak and intermediate screening, and \citet{Alastuey1978} for the strong screening regime. Plasma parameters are those from \citet{Itoh1979}. The energy-loss rates including the derivatives from thermal neutrinos are from \citet{Itoh1996}. The equation of state (EOS) is based on the OPAL EOS tables \citep{Rogers2002} including the SCVH tables \citep{Saumon1995} for lower temperatures and densities and the HELM \citep{Timmes2000} and PC \citep{Potekhin2010} EOS for higher temperatures and densities.

Convection was consider using Ledoux criterion, which takes into account the composition gradient, along with the Cox implementation of MLT \citep{Cox1968} with a mixing length free parameter $\alpha_{\rm MLT} = 2$.
The diffusion and gravitational settling in MESA are calculated by solving the equations from  \citet{Burger1969} with coefficients of \citet{Thoul1994}. In the overshoot region MESA treats convective mixing as a time-dependent diffusion process, with a diffusion coefficient given by,
\begin{equation}
D_{\rm OV} = D_{\rm conv, 0} \exp{-\frac{2z}{f \lambda_{\rm P, 0}}}
\end{equation}

\noindent where $D_{\rm conv, 0}$ is the MLT derived diffusion coefficient at the convective boundary, $z$ is the distance from the convective boundary, $\lambda_{\rm P, 0}$ is the local pressure scale height and $f$ is the adjustable parameter \citep{Herwig2000}, which can have different values at the upper and lower convective boundaries for H-burning, He-burning, metal-burning (i.e all burning regions that are not H or He-burning) and non-burning convection zones. Due to convergence issues, overshooting was only considered for the lower boundary of metal-burning convective region with parameters set to \texttt{overshoot\_f\_below\_burn\_z\_shell} $= 0.1$ and
\texttt{overshoot\_f0\_below\_burn\_z\_shell} $= 0.01$.
The mixing in regions unstable to Schwarzschild but stable to Ledoux criteria is treated by semiconvection \citep{Langer1983} using the dimensionless efficiency parameter $\alpha_{\rm sc} = 0.01$. Thermohaline mixing \citep{Ulrich1972,Kippenhahn1980} is also considered  with efficiency parameter set to $\alpha_{\rm th} = 2$. For details in the implementation of semiconvection and thermohaline mixing the reader is referred to section 4 of \citet{Paxton2013}.

Mass loss was considered during the giant phases, i.e. RGB and AGB phases. We used the mass loss formula from \citet{Reimers1975} with $\eta = 0.1$ during RGB, followed by the \citet{Bloecker1995} scheme with $\eta = 10$ on the AGB. The opacity tables are those from OPAL type 2 \citep{Iglesias1996} for enhanced C/O variations in composition. We used a gray atmosphere for the entire evolution. 
Considering white dwarf stars has rotation periods of $\sim 1$ day \citep{Kepler2017}, in most cases, the effects of rotation are negligible. Even thought there are a few known fast rotators, we do no consider rotation in our computations.

\subsection{White dwarf evolution}

Chemical diffusion and gravitational settling was modeled following the formulation from \citet{Burger1969}. \citet{Sun2017} suggested that the routine \texttt{diffusion\_use\_cgs\_solver} is better suited for electron degeneracy, however they considered only very low mass white dwarf models with stellar mass below $M_{\rm WD} < 0.17$ M$_{\rm \odot}$. Employing this formulation for diffusion in massive models leads to numerical instabilities and to non-physical chemical profiles.

In our models, convection was shut down  due to numerical instabilities during the cooling sequence.
Computations with or without convection show no differences in the final ages of our cooling sequences. The same happens when the \citet{Sun2017} formulation is considered. Thus, our treatment of diffusion and convection do not affect the cooling times in the mass range considered in this work.

The energy-loss rates from thermal neutrinos and its derivatives are from \citet{Itoh1996}.

White dwarf stars are expected to undergo crystallization as a result of strong Coulomb interactions in their dense interiors \citep{VanHorn1968}. The transition occurs when the energy of the Coulomb interaction between neighboring ions is much larger than their thermal energy. The ratio between these two contributions can be expressed as a coupling parameter $\Gamma = \bar{Z}^2 e^2/a_i k_B T$ where $\bar{Z}$ is the average ion charge, $a_i$ is the mean ion spacing and the other symbols have their usual meaning. Crystallization for pure oxygen plasma begins when $\Gamma \approx 175$ \citep{VanHorn1969}.
The onset of crystallization also depends on the adopted phase diagram. Near the crystallization limit ($\Gamma \sim 175$) MESA uses the \citet{Potekhin2010} equation of state (PC EOS) which accounts for the thermodynamics of crystals. By default, MESA changes from HELM EOS to PC EOS when the Coulomb coupling parameter $\Gamma > 80$ and considers a mixture of solid and liquid for $\Gamma_{\rm i} = 150$ and a full crystal when $\Gamma_{\rm full} = 175$.
Those values where obtained by \citet{Potekhin2010}. Results from the asteroseismological analysis performed by \citet{Romero2013} indicates that the azeotropic type phase diagram from \citet{Horowitz2010} \citep[see also][]{Schneider2011, Hughto2012} better represents the crystallization on white dwarfs cores. Hence, we modified the values for the coupling constant $\Gamma$ to $\Gamma_{\rm i} = 215$ and $\Gamma_{\rm full} = 220$ in our computations. Thus, the release of latent heat occur for $\Gamma = 215-220$. This modification had to be done on the file \texttt{pc\_eos.f} located inside MESA \texttt{eos} module. Until MESA version r8845 used in this work, there was no option to control those parameters. 

The Debye cooling regime, which affects the cooling of white dwarfs at lower luminosities, was not considered in our simulations. However, this effect will not be important in our results, since it occurs for temperatures below $\sim 10\, 000$ K and luminosities lower than $\log L/L_{\rm \odot} < -4.06$ \citep{Althaus2007} for the mass range considered in this work (see section \ref{cooling}).

\section{Evolution from the ZAMS to the cooling curve}
\label{sec:3-mass_evolution_before_cs}

We calculated 16 full evolutionary sequences with initial metallicity $Z=0.02$ and initial mass at the ZAMS between 8.80 and 11.80 M$_{\rm \odot}$. As a result we obtained white dwarf models with stellar masses ranging from $1.012$ to $1.307$ M$_{\rm \odot}$. From our computations, we find that a sequence with an initial mass of $11.9$ M$_{\rm \odot}$ do not result in a white dwarf but possibly ends up as a neutron star. Thus, a white dwarf with the Chandrasekhar limiting mass should have a progenitor with an initial mass between $11.8$ and $11.9$ M$_{\rm \odot}$.
Further
calculations are needed to determine the model which lead to the Chandrasekhar mass limit. Note that within this mass range there are two possible core compositions, i.e, C/O and O/Ne/Mg. Our computations complement at high stellar masses the works of \citet{Romero2012, Romero2013} who computed C/O core white dwarf sequences with masses below $M_{\rm WD} = 1.050$ M$_{\rm \odot}$.
Our H and He atmosphere sequences have evolved equally on the stages prior to the cooling sequence, therefore there is no distinction on the evolution previous to the white dwarf stage.

\begin{figure}
	\includegraphics[width=\columnwidth]{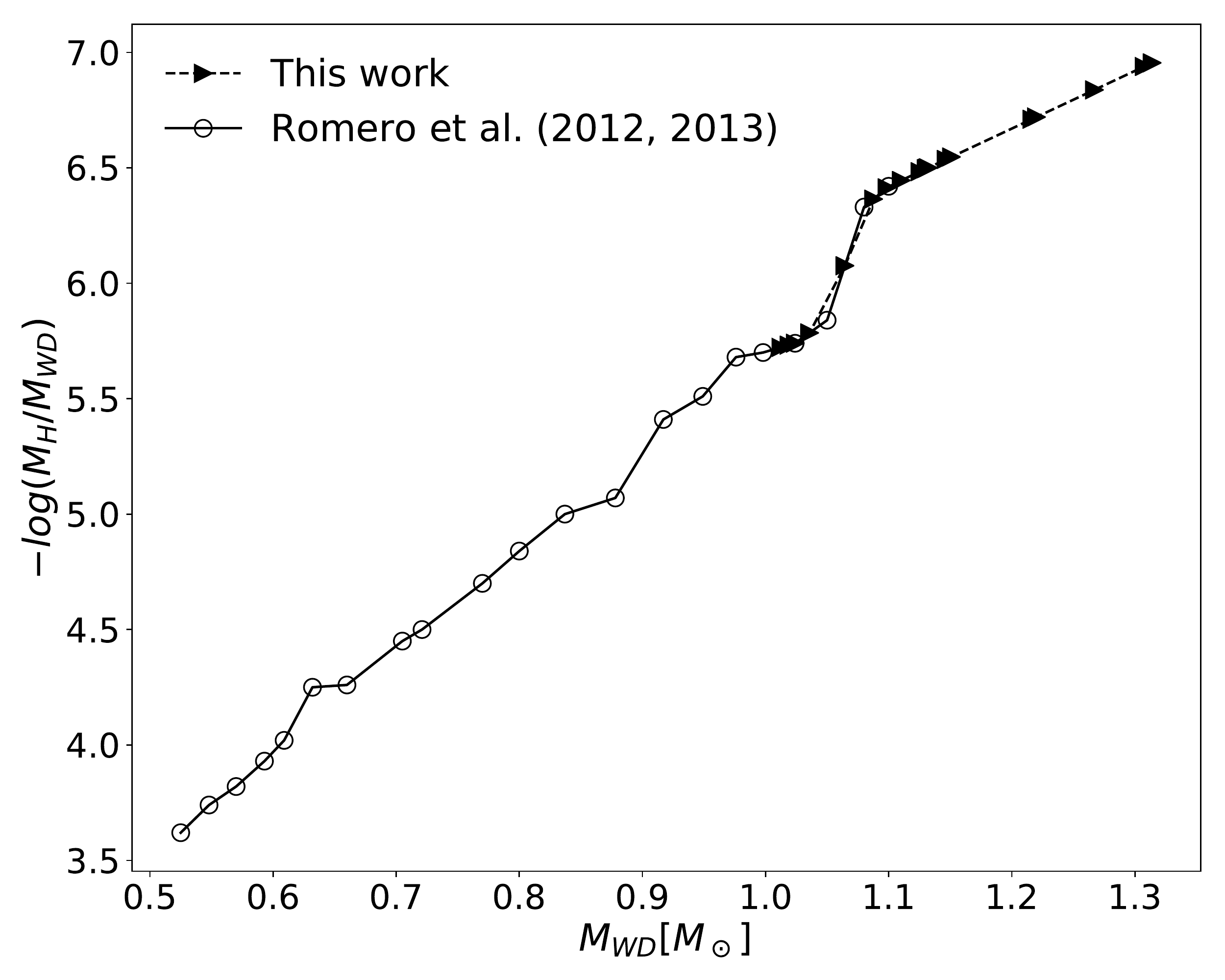}
    \caption{Range of star mass and hydrogen envelope mass. Circles are from \citet{Romero2012,Romero2013} and triangles are the extrapolated data.}
    \label{fig:h_envelope}
\end{figure}

During the AGB, the computations present numerical difficulties related to the mass loss episodes. At this stage, we use the mass loss prescription of \citet{Bloecker1995} with a factor $\eta$=10, that leads to the complete removal of the hydrogen envelope, preventing the occurrence of thermal pulses, in all sequences. A less efficient mass loss rate, achieve by decreasing the value of $\eta$, does not allow us to compute the stages following the AGB stage but still leads to the complete removal of hydrogen. The same occurs for other MESA built in mass loss schemes.
\citet{Garcia-berro97-2} discussed that models with rapid radiative wind during AGB phase are expected to lose all its hydrogen layer and most of its helium layer leading to a PG 1159 star, but real massive white dwarfs are known with H and He envelopes. In order to produce hydrogen atmosphere white dwarf sequences, we artificially added hydrogen at the surface of the models. This procedure was performed at high effective temperature, near the beginning of the cooling sequences, so the transitory effects caused by the artificial procedure are rapidly washed out. The amount of hydrogen added to each sequence was computed by a linear extrapolation to higher stellar mass pf the results of \citet{Romero2012,Romero2013}. The amount of hydrogen as a function of the white dwarf stellar mass is shown in Figure \ref{fig:h_envelope}. Circles correspond to the results from \citet{Romero2012,Romero2013}, while the triangles correspond to the values extrapolated for higher stellar masses.

\begin{figure}
	\centering
	\includegraphics[width=\columnwidth]{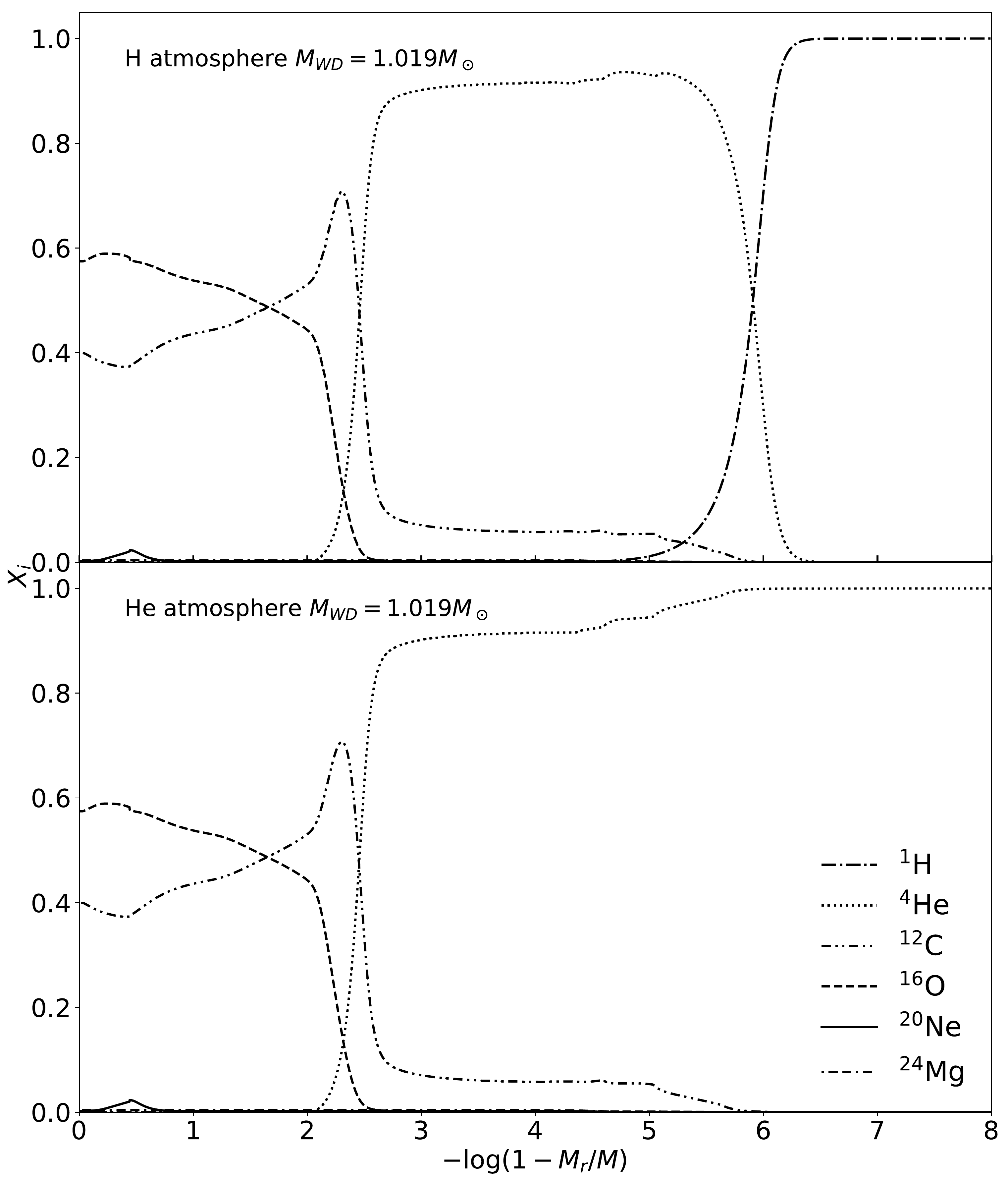}
	\caption{Example of chemical profile in terms of the outer mass fraction at $T_{\rm eff} \approx 80,000$ K on the cooling sequence for H atmosphere (top) and He atmosphere (bottom) white dwarf with 1.019 M$_{\rm \odot}$. Hydrogen had to be added on all sequences due to losses on mass loss phase. The quantities added on each evolution are shown in table~\ref{tab:envelopes}. This figure show that only the atmosphere of the star was changed. The nuclei composition are identical to both H and He atmosphere sequences.}
	\label{fig:cs_initial_composition}
\end{figure}

As an example, Figure \ref{fig:cs_initial_composition} shows two chemical profile in terms of outer mass fraction at $T_{\rm eff} \approx 80\,000$ K on the cooling curve for a sequence with stellar mass $M_{\rm WD} = 1.019$ M$_{\rm \odot}$ and a C/O core. The bottom panel shows the chemical profile for the helium atmosphere case and the top panel presents the chemical profile for the hydrogen atmosphere sequence, after $M_{\rm H} = 10^{-5.7} M_{\rm WD}$ of hydrogen was artificially added to the model. Note that the only difference between the two chemical profiles shown in Figure  \ref{fig:cs_initial_composition} is the chemical abundance of the outer layers, while the core regions remain the same.

\begin{figure}
	\includegraphics[width=\columnwidth]{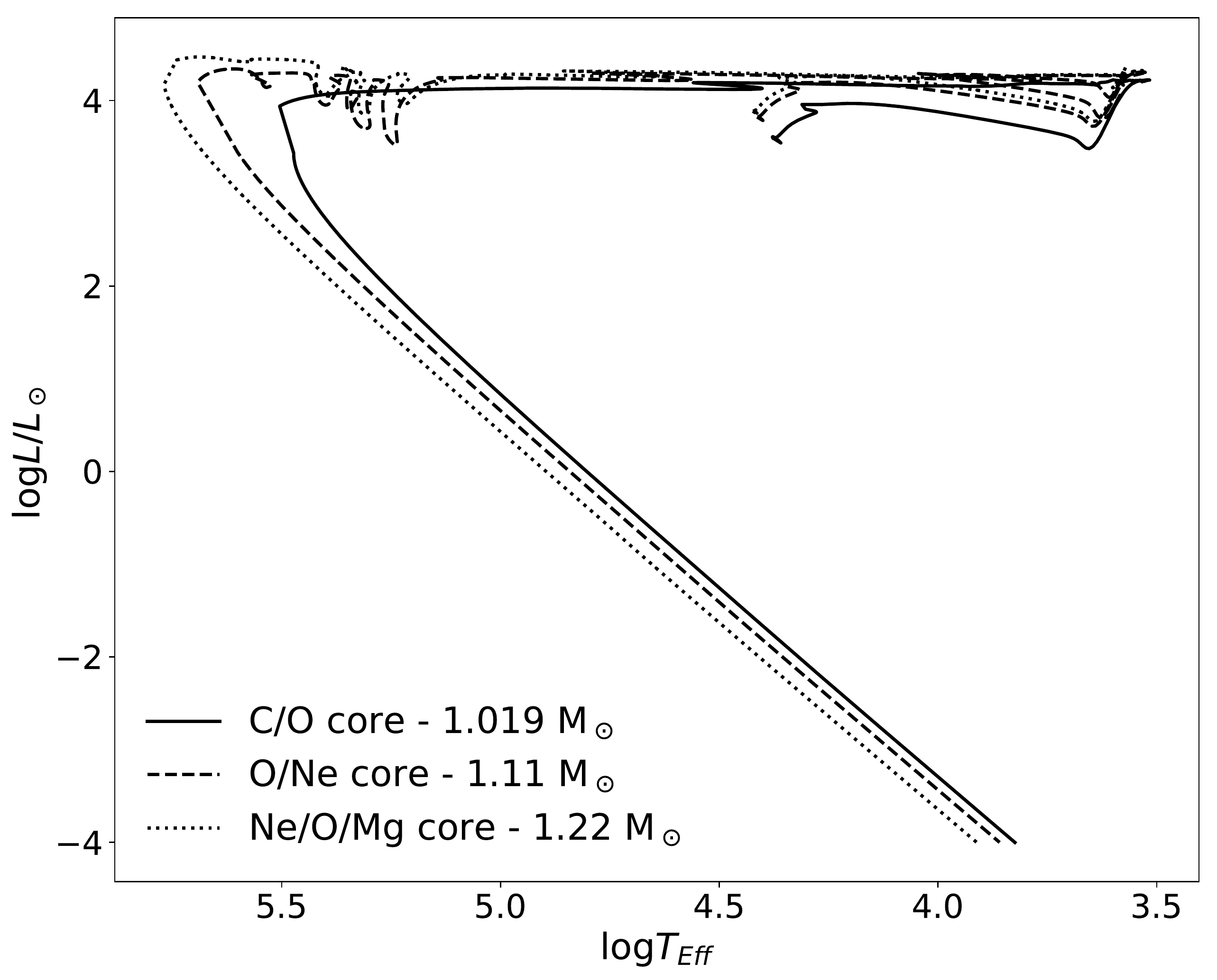}
    \caption{Evolutionary tracks for three sequences in the HR diagram, with initial masses 8.9 (solid line), 10 (dashed line) and 11 M$_{\rm \odot}$ (dotted line) and final masses 1.019, 1.11 and 1.22 M$_{\rm \odot}$, respectively. Each sequence has a different core composition in the white dwarf cooling curve stage, being C/O, O/Ne and Ne/O/Mg for the sequences with white dwarf masses 1.019, 1.11 and 1.22 M$_{\rm \odot}$, respectively.}
    \label{fig:hrds_cores}
\end{figure}

The sequences with final masses $M_{\rm WD} \ge 1.024$ M$_{\rm \odot}$ experience loops at high luminosity and high effective temperature region in the post-AGB stage (see Figure \ref{fig:hrds_cores}). The loops are due to carbon shell-burning followed by helium shell-burning which, for sequences with $M_{\rm WD} \ge 1.132$ M$_{\rm \odot}$, burns all the helium content in the outer layers. These helium shell-burning happens due to the huge temperature gradient on the shells surrounding the core. Thus, enough shells must be added in the numerical models in these regions to account for the large temperature gradients and composition gradients in the shell burning regions.

For sequences with $M_{\rm WD} \ge 1.132$ M$_{\rm \odot}$ helium was added on top of the model at the beginning of the cooling sequence using the same procedure used to add hydrogen. The amount of helium to be added was extrapolated from our sequences which had not burnt all helium. The amount of hydrogen and helium for each sequence is presented in columns 3 and 4 of Table \ref{tab:envelopes}, along with the initial mass at the ZAMS and the white dwarf mass in column 1 and  2, respectively. Helium atmosphere sequences were produced by not adding hydrogen on top of the models.

\begin{table}
    \centering
    \caption{Values of initial mass at ZAMS ($M_{\rm ZAMS}$), White dwarfs masses ($M_{\rm WD}$) and hydrogen content for H-rich sequences and helium contents for both H and He atmosphere sequences. Hydrogen abundances where extrapolated from \citet{Romero2012,Romero2013}. The Helium abundances were extrapolated for sequences with $M_{\rm WD} \ge 1.132$.}
    \label{tab:envelopes}
    \begin{tabular}{cccc}
        \hline
        $M_{\rm ZAMS}$ $[M_{\rm \odot}]$ & $M_{\rm WD}$ $[M_{\rm \odot}]$ & $-\log M_{\rm H}/M_{\rm WD}$ & $-\log M_{\rm He}/M_{\rm WD}$ \\
        \hline
          8.80 &  1.012 &  5.722 &      2.451 \\
          8.90 &  1.019 &  5.732 &      2.466 \\
          9.00 &  1.024 &  5.740 &      3.477 \\
          9.20 &  1.036 &  5.785 &      3.349 \\
          9.50 &  1.064 &  6.076 &      2.958 \\
          9.80 &  1.088 &  6.365 &      4.006 \\
         10.00 &  1.110 &  6.446 &      3.517 \\
         10.10 &  1.125 &  6.484 &      2.447 \\
         10.20 &  1.131 &  6.498 &      2.489 \\
         10.30 &  1.132 &  6.500 &      2.495 \\
         10.45 &  1.147 &  6.536 &      2.597 \\
         10.50 &  1.151 &  6.547 &      2.628 \\
         10.80 &  1.216 &  6.711 &      3.089 \\
         11.00 &  1.220 &  6.720 &      3.116 \\
         11.40 &  1.267 &  6.838 &      3.449 \\
         11.80 &  1.307 &  6.939 &      3.735 \\
        \hline
        \end{tabular}
    \end{table}

The Hertzprung-Russell diagram from ZAMS to white dwarf cooling sequence is shown in Figure \ref{fig:hrds_cores} for three sequences with initial masses 8.9, 10 and 11 M$_{\rm \odot}$ and final masses of 1.019, 1.11 and 1.22 M$_{\rm \odot}$ in the cooling curve, respectively. Each sequence in this figure has a different core composition as a white dwarf star. The solid, dashed, and dotted lines represent a C/O core white dwarf with $M_{\rm WD} = 1.019$ M$_{\rm \odot}$, an O/Ne core white dwarf with $M_{\rm WD} = 1.11$ M$_{\rm \odot}$ and a Ne/O/Mg white dwarf with $M_{\rm WD} = 1.22$ M$_{\rm \odot}$, respectively.

The $M_{\rm WD} = 1.019$ M$_{\rm \odot}$ sequence with C/O core experienced a core hydrogen burning stage lasting for $24.51$ Myr, and then during the next $2.80$ Myr helium has burn until depletion at the center. For more massive sequences the time spent on main sequence and central helium burning decreases. For the sequences with final mass $M_{\rm WD} = 1.11$ M$_{\rm \odot}$, with a O/Ne core, the total main sequence time is $19.75$ Myr plus $2.28$ Myr to burn all helium. Lastly, the sequence with final mass $M_{\rm WD} = 1.22$ M$_{\rm \odot}$ burnt hydrogen for $16.83$ Myr until depletion at the center and burnt helium for $1.81$ Myr until extinction at the center. The ages for hydrogen and helium depletion at center and progenitor age, i.e., the age at the beginning of the white dwarf cooling sequence, are listed in columns 3, 4 and 5 of Table \ref{tab:ages_teff_cristal}, respectively (see section \ref{cooling}).

\subsection{Carbon flame}
\label{sec:carbon_flame}

As can be seen from Figure \ref{fig:hrds_cores}, the sequences characterized with white dwarf masses of $1.11 M_{\rm \odot}$ and $1.22 M_{\rm \odot}$ presents loops on the upper left region of the HR diagram, prior to entering the white dwarf cooling sequence. Those loops are due to carbon burning which ignites off-center and moves inwards to the center of the star. In particular, towards the end of the loop, just before the extinction of the carbon burning, helium ignites on a shell near the surface.
All sequences with $M_{\rm WD} \geq 1.024$ M$_{\rm \odot}$ experiences an off-center carbon ignition prior to the white dwarf stage, and for sequences with $M_{\rm WD} \geq 1.132$ M$_{\rm \odot}$ the subsequent helium burning consumes all the helium content in the outer layers.
If the carbon flame do not reach the center of the star, a hybrid C/O-O/Ne white dwarf will form \citep{Denissenkov2013}, while if the flame propagates to the center a O/Ne/Mg white dwarf will be produced. The occurrence of the late carbon flashes can be an explanation for the luminosity variations observed in the star in the Stingray Nebula, CD-59 6479  or SAO 24567 \citep{Henize1976,Reindl2014}. The luminosity variation timescale at the beginning of the carbon burning in our sequence with $M_{\rm WD} = 1.024$ M$_{\rm \odot}$ are similar with the one observed, however, the total change in luminosity in CD-59 6479   is larger than the one showed in our models by a factor of $\sim 100$ \citep{Schaefer2015} and the stellar masses are very different.

\begin{figure}
	\includegraphics[width=\columnwidth]{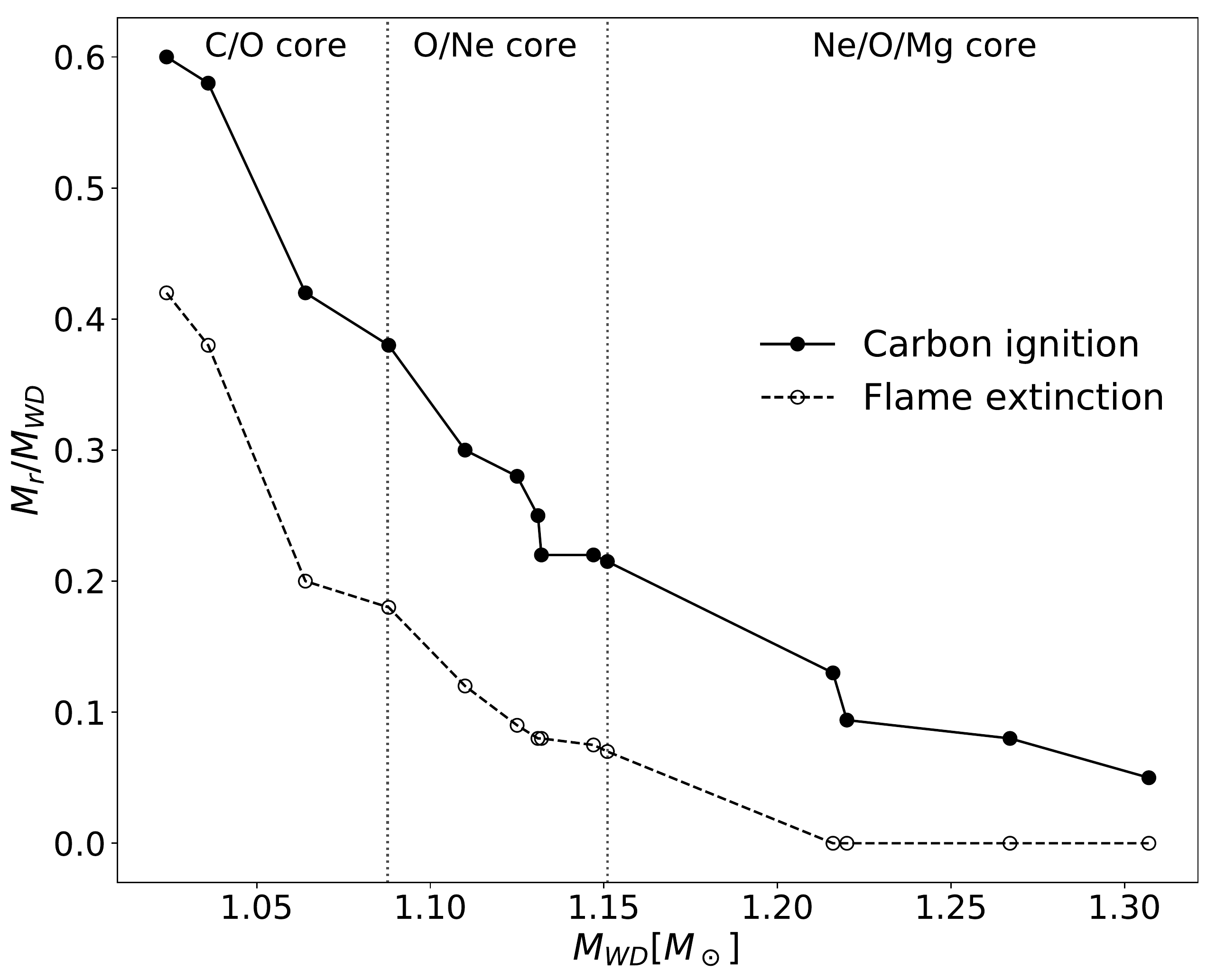}
    \caption{Mass location of the first ignition of carbon and mass location of closest to center flame extinction as function of white dwarf masses which experienced a carbon flame. Vertical dotted lines delimits the mass range for core composition. See text for details.}
    \label{fig:carbon_flame}
\end{figure}

All sequences which experienced a carbon flame presented subsequent flames which started further away from the center and were quenched before the starting point of the previous flame.
Figure \ref{fig:carbon_flame} shows the location in mass where the first carbon ignition occurs (full line) and the location of the extinction of the carbon flame which got closest to center (dashed line), as a function of the white dwarf mass which experienced a carbon flame. Note that the sequences with stellar mass $M_{\rm WD} < 1.024$ M$_{\rm \odot}$ do not experience carbon ignition. The dotted vertical lines separate the mass range for different core compositions (see subsection \ref{subsec:central_abun}). As can be seen from this figure, the position of the first  carbon ignition gets closer to the center of the model for increasing white dwarf mass. As the core mass increases, the central degeneracy is lower, allowing the inner maximum temperature to be near the center, causing a carbon ignition closer to center on more massive sequences.

Specifically, the sequence with $M_{\rm ZAMS} = 10$ M$_{\rm \odot}$ ($M_{\rm WD} = 1.11$ M$_{\rm \odot}$) experience an off-center carbon ignition starting at $M_{\rm r}/M_{\rm WD} \sim 0.3$ which moves inwards until $M_{\rm r}/M_{\rm WD}  \sim 0.1$, while for the sequence with $M_{\rm ZAMS} = 11.0$ M$_{\rm \odot}$ ($M_{\rm WD} = 1.220$ M$_{\rm \odot}$) carbon ignition starts at $M_{\rm r}/M_{\rm WD}  \sim 0.094$ and, in this case, reaches the center of the star.
Carbon ignition in all sequences started on regions with density $\rho \sim 1.5\times10^6$  g/cm$^{3}$, agreeing with the results for super-AGB sequences from \citet{Farmer2015}. For a detailed discussion about carbon burning we refer to the works of \citet{Siess2006,Siess2007} and of \citet{Farmer2015} and references therein.

From our computations, some sequences with C/O core in the white dwarf stages experienced a carbon ignition, but the carbon flame only reaches $M_{\rm r}/M_{\rm WD} \sim 0.2$ for those models. When the carbon flame reaches the regions closer than $M_{\rm r}/M_{\rm WD} < 0.2$, a hybrid C/O-O/Ne white dwarf with O/Ne core is formed. For sequences with $M_{\rm WD} > 1.15$ M$_{\rm \odot}$ the carbon flame reaches the center of the star producing a Ne/O/Mg core white dwarf.
The details of core composition and chemical profile will be analyzed on the following subsections.

\subsection{The Initial-to-Final Mass Relation}
\label{sec:6-ifmr}

\begin{figure}
    \centering
    \includegraphics[width=\columnwidth]{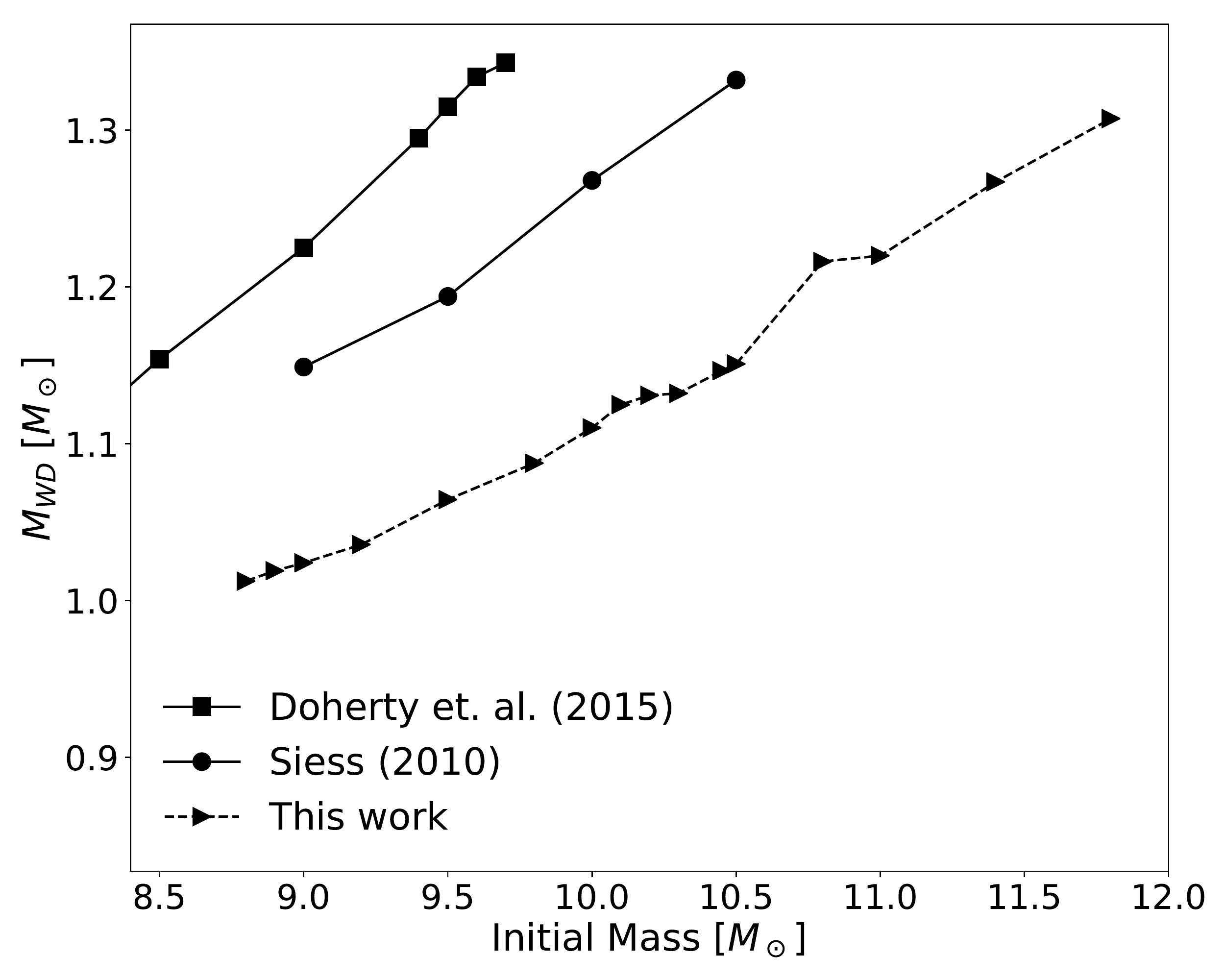}
    \caption{Initial-to-final mass relation obtained from our computations (triangle)  as compared to the results from \citet{Siess2010} (circle) and \citet{Doherty2015} (square). In all cases the initial metallicity is $Z=0.02$. We only show the massive models of \citet{Doherty2015}.}
    \label{fig:ifmr}
\end{figure}

Figure \ref{fig:ifmr} presents the initial-to-final mass relation obtained from our computations compared to the results of \citet{Siess2010} and \citet{Doherty2015}, within the stellar mass range considered in this work. We consider models with metallicity $Z=0.02$. The final mass obtained in this work is systematically lower than previous results, due to the different input physics considered in each case. \citet{Siess2010} considers the \citet{Vassiliadis1993} mass loss scheme, a mixing length parameter of $\alpha_{\rm MLT} = 1.75$ and no core overshooting. On the other hand, \citet{Doherty2015} consider mass loss rate from \citet{Reimers1975} on the RGB, followed by \citet{Bloecker1995} at high luminosities and change to \citet{Vassiliadis1993} on thermally pulsing phase, with similar convection treatment than \citet{Siess2010}. Also, both authors evolved their sequences only until the TP-AGB phase, and then consider the mass of the He-free core as the final mass at the white dwarf stage.
As discussed in \citet{Siess2010} the final mass is dependent on core growth and mass loss scheme.
In our computations, we calculated full evolutionary sequences by employing the mass loss formula from \citet{Reimers1975} on RGB and \citet{Bloecker1995} on the AGB with a larger efficiency parameter than the mass loss rates considered by previous authors, thus evolving the sequences to post-AGB phase and leading to a lower final mass in the cooling curve for the same initial mass. Thus, from our computations, the upper limit of progenitor mass to produce a white dwarf star is $M_{ZAMS} = 11.8$ M$_{\rm \odot}$ for $Z=0.02$, $ 21.6 \%$ greater than the value found by \citet{Doherty2015} and $12.4 \%$ greater compared to \citet{Siess2010} for the same initial metallicity. Considering the dispersion in the initial-to final Mass relation from the different authors in figure \ref{fig:ifmr}, we can estimate for a given initial mass in the ZAMS a range of final white dwarf masses. For instance, a sequence with initial mass of $9$ M$_{\rm \odot}$ can produce a white dwarf with masses ranging from $1.024$ to $1.225$ M$_{\rm \odot}$, depending on the input physics applied, specially the mass loss formulation. The present day observational data cannot distinguish the best model.

\subsection{Central abundances in the cooling sequence}
\label{subsec:central_abun}

\begin{figure}
	\includegraphics[width=\columnwidth]{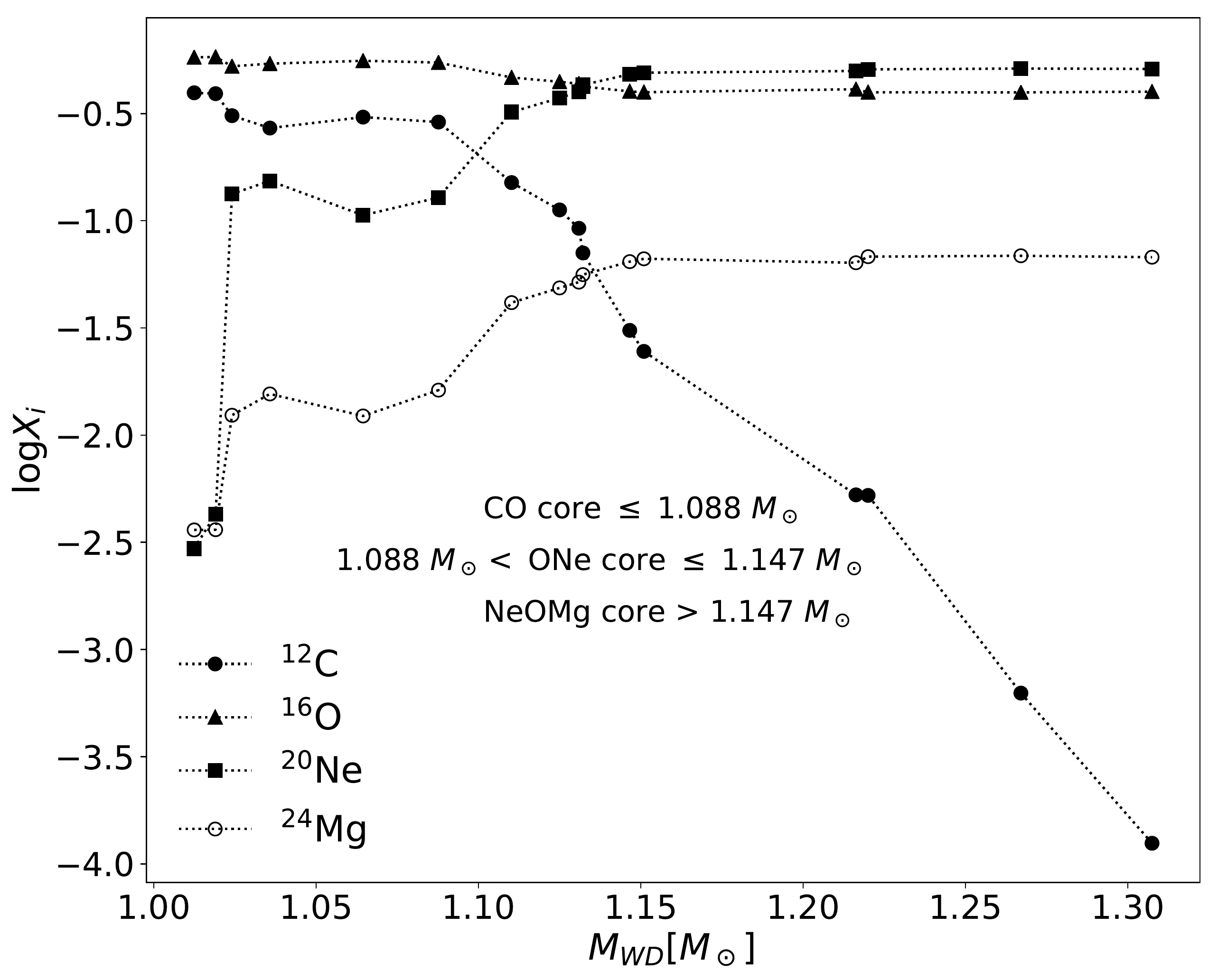}
    \caption{Central abundances versus final mass of the white dwarfs stars before crystallization starts ($T_{\rm eff} \approx 40\, 000$ K). The black circles represent $^{12}$C, triangles are $^{16}$O, squares are $^{20}$Ne and empty circles are $^{24}$Mg. We can identify three regions, one for C/O core for masses below or equal to $1.088$ M$_{\rm \odot}$ where neon is lower than carbon, another region between $1.088$ and $1.147$ M$_{\rm \odot}$ in which neon is greater than carbon and lower than oxygen but magnesium is lower than carbon defining a O/Ne core, and a third region for masses greater than $1.147$ M$_{\rm \odot}$ with a O/Ne/Mg core due to neon being greater than oxygen and magnesium greater than carbon.}
    \label{fig:central_abundancies}
\end{figure}

Within the stellar mass range studied in this work, we expect to have three main core compositions, i.e., C/O and O/Ne \citep{Garcia-berro1997,Doherty2015} or Ne/O/Mg, changing with increasing stellar mass. Figure \ref{fig:central_abundancies} shows central abundances of carbon, oxygen, neon and magnesium in a logarithmic scale, as a function of the white dwarf mass.
The values are taken at
$T_{\rm eff} \approx 40\, 000$ K, a effective temperature low enough to give time for diffusion to act on the central composition and still high enough to be above crystallization. As shown on Figure \ref{fig:central_abundancies}, the C/O core is dominant for $M_{\rm WD} \leq 1.088$ M$_{\rm \odot}$ due to carbon being more abundant than neon. Between $1.088 < M_{\rm WD} \leq 1.147$ M$_{\rm \odot}$ we have an O/Ne core as the amount of neon increases but is still lower than oxygen, whilst for masses $M_{\rm WD} > 1.147$ M$_{\rm \odot}$ neon is the most abundant element on the central composition, followed by oxygen and then by magnesium, defining the core composition as Ne/O/Mg core. Note that sequences with $M_{\rm WD} > 1.15$ M$_{\rm \odot}$ experienced a carbon flame which reaches the center of the stars as shown on Figure \ref{fig:carbon_flame} and discussed on subsection \ref{sec:carbon_flame}. Also, the sequence with $M_{\rm WD} = 1.024$ M$_{\rm \odot}$ is the sequences with the lowest stellar mass to experience carbon burning, but it does not reach the center, producing an hybrid C/O-O/Ne white dwarf. This explains the step in the central $^{20}$Ne abundance observed in Figure \ref{fig:central_abundancies} for that stellar mass.
\citet{Doherty2015} computed sequences for super-AGB stars and found that, for $Z=0.02$, sequences with core masses $M_{\rm c} < 1.075$ M$_{\rm \odot}$ become a C/O core white dwarf, sequences with core masses $ 1.075 \leq M_{\rm c} < 1.154$ M$_{\rm \odot}$ presented a C/O/Ne core and for core masses $M_{\rm c} \geq 1.154$ M$_{\rm \odot}$ the sequences are O/Ne white dwarfs. \citet{Woosley2015} also calculated sequences on the limit of SN progenitors and obtained models with O/Ne core for a range of core mass $1.088 \leq M_{\rm c} \leq 1.345$ M$_{\rm \odot}$. Chemical profiles from \citet{Doherty2015} and \citet{Woosley2015} are not publicly available so we can not compare the internal abundances of the last models computed.
It is important to note that \citet{Doherty2015} and \citet{Woosley2015} did not compute the cooling curves, opposite to the present work. They also uses different evolutionary codes with different nuclear reaction and input physics.

\subsection{Chemical profiles in the white dwarf cooling sequence}

\begin{figure}
	\includegraphics[width=\columnwidth]{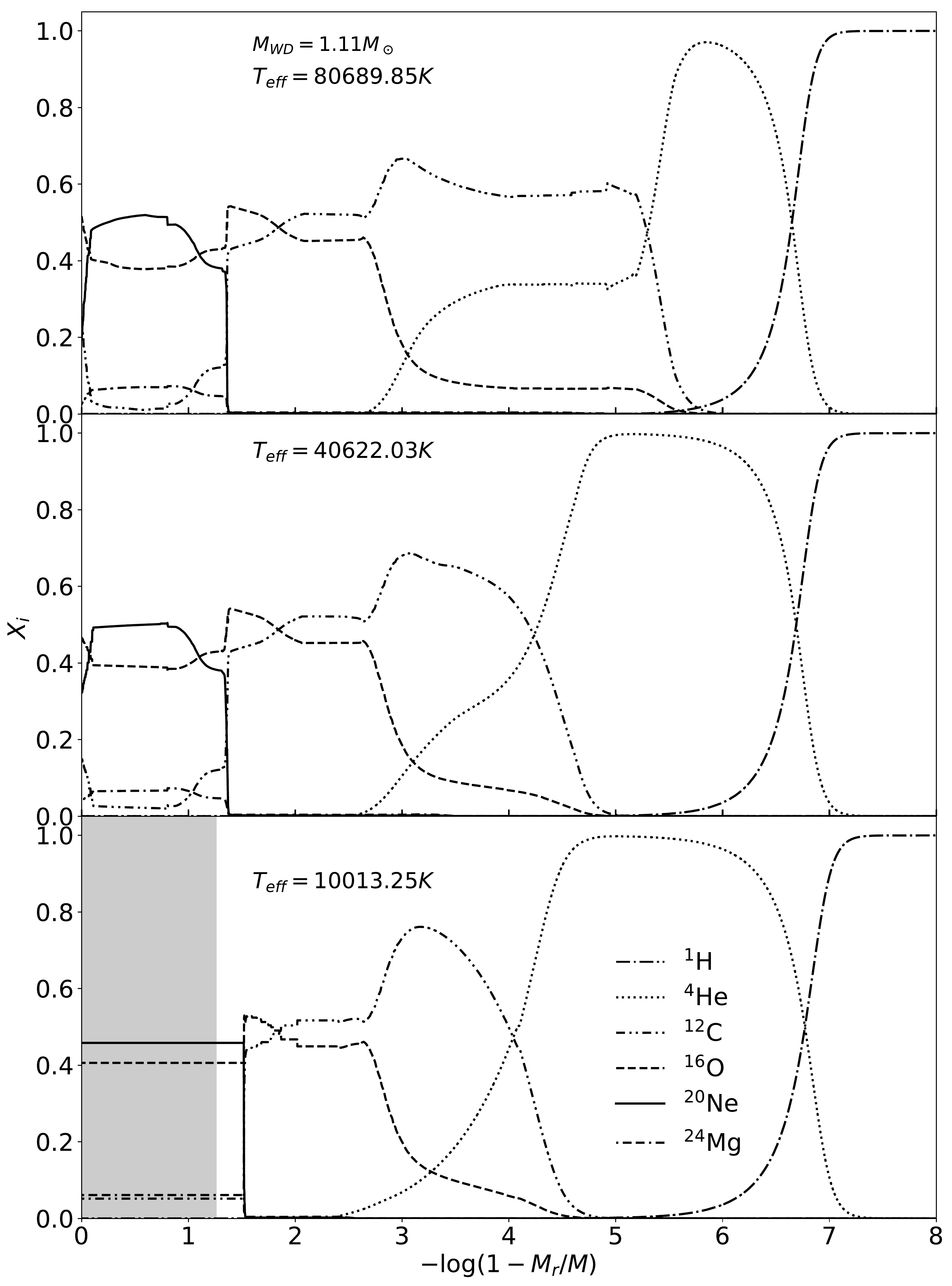}
    \caption{Chemical profile in terms of outer mass fraction for selected elements for three different stages of the cooling sequence: $T_{\rm eff} \approx 80\,000$ K on top, $T_{\rm eff} \approx 40\,000$ K on middle and $T_{\rm eff} \approx 10\,000$ K on bottom panels. The chemical profiles are for a H atmosphere white dwarf with $M_{\rm WD} = 1.11 M_{\rm \odot}$. The shaded region represents the crystallized zone of the star.}
    \label{fig:abun_along_cs}
\end{figure}

In Figure \ref{fig:abun_along_cs} we show three chemical profiles for models along the white dwarf cooling sequence for an O/Ne core H--atmosphere white dwarf with stellar mass $M_{\rm WD} = 1.11$ M$_{\rm \odot}$. The abundance profiles are in terms of outer mass fraction and correspond to three effective temperatures, $T_{\rm eff} \approx 80\, 000 , 40\, 000$ and $10\, 000$ K, on top, middle and bottom panels, respectively. The effects of diffusion and gravitational settling on the entire chemical profile as the star cools are evident from this figure. Initially, the oxygen central abundance is $X_{\rm O} = 0.515$ while for neon it is $X_{\rm Ne} = 0.195$ (top panel). Rehomogenization processes changes the central abundances to a flat profile with $X_{\rm O} = 0.406$ and $X_{\rm Ne} = 0.458$ when the star reaches  effective temperatures $\sim 10, 000$ K (bottom panel). Also there is a carbon-oxygen-helium triple-layer structure at $2.7 < - \log(1-M_{\rm r}/M) < 5.5$ for high effective temperatures, which diminish due to diffusion processes.
 Note that, for the sequence with white dwarf mass $1.11$ M$_{\odot}$ the carbon flame experienced in the pre-white dwarf stages, did not reach the center of the model, giving rise to a  hybrid C/O-O/Ne core white dwarfs \citep{Denissenkov2013,Doherty2015,Farmer2015}. The hybrid models presented in \citet{Denissenkov2013} are characterized by a C/O core surrounded by an O/Ne region at the super AGB stage. This chemical structure differs from the one characterizing our models (see Figure \ref{fig:abun_along_cs}) , since ours are taken from the white dwarf cooling curve, where rehomogenization process are taken into account.
To our knowledge this is the first time a chemical profile for the entire star is presented for a hybrid C/O-O/Ne white dwarf. The shaded region on the bottom panel depicts the crystallized core, which is defined as the  regions where $\Gamma \geq 220$ (see section \ref{crist}). For this sequence, crystallization starts at $T_{\rm eff} \approx 20\, 000$ K. Finally, the chemical structure of the star can be separated into the outer layers, composed by hydrogen followed by an extended helium buffer, and the inner regions showing an triple-layer of carbon-oxygen-helium from $2 < - \log(1-M_{\rm r}/M) < 4$ with a carbon peak ($-\log M_{\rm r}/M \sim 3.1$) on top of the Ne/O core.
This chemical structure of the core is the result of burning events occurring on previous stages to the white dwarf cooling sequence.

\begin{figure}
	\includegraphics[width=\columnwidth]{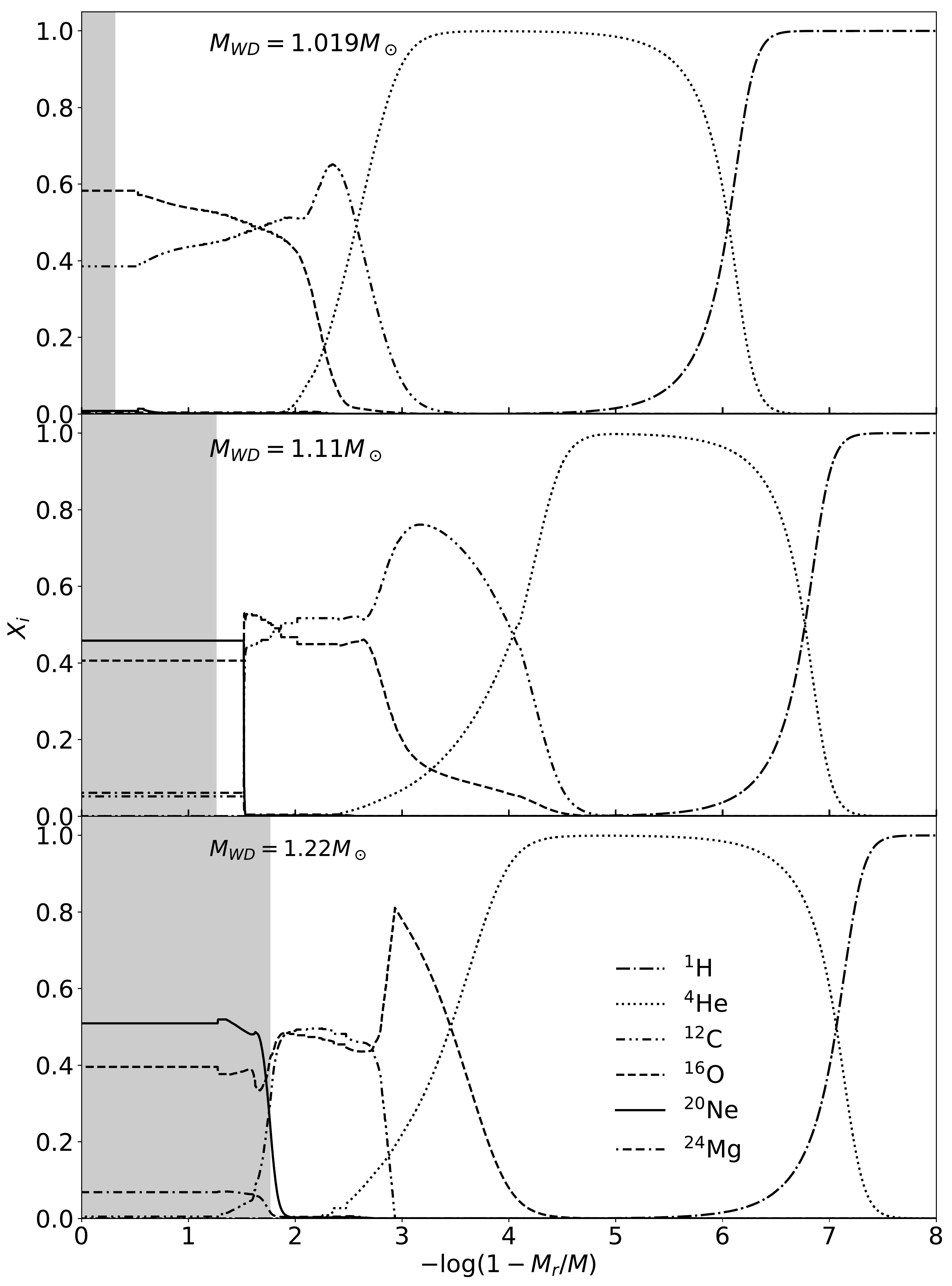}
    \caption{Chemical profile in terms of outer mass fraction for selected elements and for three different H atmosphere white dwarf with $T_{\rm eff} \approx 10\, 000$ K. From top to bottom we depict the chemical profile for a $M_{\rm WD} = 1.019, 1.11, 1.22$ M$_{\rm \odot}$. The shaded region represents the crystallized zone of the star}
    \label{fig:comp_chemical_profile_da}
\end{figure}

The composition of the core and its size has a strong impact on the chemical profile of the entire star.
Figure \ref{fig:comp_chemical_profile_da} shows the chemical profiles in terms of outer mass fraction for three hydrogen atmosphere white dwarfs models with $T_{\rm eff} \approx 10\, 000$ K, with stellar mass $M_{\rm WD} = 1.019$ M$_{\rm \odot}$ (top panel), $M_{\rm WD}=1.11$ M$_{\rm \odot}$ (middle panel) and  $M_{\rm WD} =1.22$ M$_{\rm \odot}$ (bottom panel). The shaded region represents the crystallized region of the model. From this figure, we note that the crystallization front is closer to the surface of the model for higher stellar mass, located at $-\log (1-M_{\rm r}/M) \approx 0.9, 1.2$ and $2$ for $M_{\rm WD} = 1.019$, $M_{\rm WD} = 1.11$ and $M_{\rm WD} = 1.22$  M$_{\rm \odot}$ models, respectively.
The sequence characterized by a stellar mass $M_{\rm WD} = 1.019$ M$_{\rm \odot}$ has a C/O core with a core mass of $M_{\rm c} = 1.016$ M$_{\rm \odot}$ extending to $-\log (1-M_{\rm r}/M_*) = 2.6$ and present a triple-layer of carbon-oxygen-helium at $1.8 < -\log (1-M_{\rm r}/M_*) < 2.6$. On top of the core, the model has a helium buffer of $M_{\rm He}= 10^{-2.47} M_{\rm WD}$ and a hydrogen envelope with $M_{\rm H} =10^{-5.73} M_{\rm WD}$. All these structures have strong impact on the propagation of stellar oscillations, when present \citep{Winget1997}.
\citet{Althaus2010} and \citet{Romero2012,Romero2013}  also obtained a triple-layer structure in their C/O core white dwarf models.

The sequence with $M_{\rm WD} = 1.11$ M$_{\rm \odot}$ has an O/Ne core, extending to $-\log (1-M_{\rm r}/M) = 4$,  which represents in mass $M_{\rm c} = 1.1099$ M$_{\rm \odot}$. As discussed before, this O/Ne core model shows a triple-layer oxygen-carbon-helium with a peak of carbon at the outer boundary of the core, a helium buffer of $M_{\rm He} = 10^{-3.52} M_{\rm WD}$ and a hydrogen envelope of $M_{\rm H} = 10^{-6.45} M_{\rm WD}$.  The $M_{\rm WD} = 1.22$  M$_{\rm \odot}$ sequence has a Ne/O/Mg core with $M_{\rm c} = 1.219$ M$_{\rm \odot}$ which extends to $-\log (1-M_{\rm r}/M) = 3.5$. It also presents a triple-layer of oxygen-carbon-helium with a sharp oxygen peak, instead of a carbon peak in the O/Ne core case. The Ne/O/Mg core model has an helium buffer of $M_{\rm He} = 10^{-3.12} M_{\rm WD}$ and a hydrogen envelope of $M_{\rm H} =  10^{-6.72} M_{\rm WD}$.
The difference in core composition are due to final mass as shown on Figure  \ref{fig:central_abundancies}. Asteroseismology will be able to distinguish these structures.

In Appendix \ref{sec:appendix-1} we present all chemical profiles for both H and He atmosphere sequences at  effective temperatures $T_{\rm eff} \approx 10\,000$ K.

\subsection{Crystallization and the Coulomb coupling parameter}
\label{crist}

As stated before, the MESA code considers as a standard input value that a mixture of solid and liquid phases occurs for a coupling parameter $\Gamma_{\rm i} = 150$ while a full crystal structure occurs when $\Gamma_{\rm full} = 175$. However, \citet{Romero2013} calculated white dwarfs sequences considering two types of phase diagram and concluded that the azeotropic type from \citet{Horowitz2010} better represents the crystallization on the nuclei of white dwarfs. Hence, we modify the values of the $\Gamma$ parameter to $\Gamma_{\rm i} = 215$ for a liquid/solid coexisting phases and $\Gamma_{\rm full} = 220$ for a full crystal phase, in line with the results obtained using the \citet{Horowitz2010} phase diagram. \citet{Paxton2017} calculated cooling sequences for a $0.6$ M$_{\rm \odot}$ white dwarf star, using MESA, varying $\Gamma_{\rm i} = 174 \ \text{to} \ 220$ and $\Gamma_{\rm full} = 176 \ \text{to} \ 240$ and found a difference in the cooling times of the order of $\sim 0.2$ Gyr. For the stellar mass range considered in this work, to assume $\Gamma_{\rm i} = 215$ and  $\Gamma_{\rm full} = 220$, makes the effective temperature at which crystallization starts $\approx 2000$ K lower, and the age $\approx 0.23$ Gyr larger than the ones using the standard input values.
The effective temperature and age on the onset of crystallization for three sequences considering the two values of $\Gamma_{\rm full}$, are listed in table \ref{tab:diff_in_age_for_gammas}.

\begin{figure}
    \centering
    \includegraphics[width=\columnwidth]{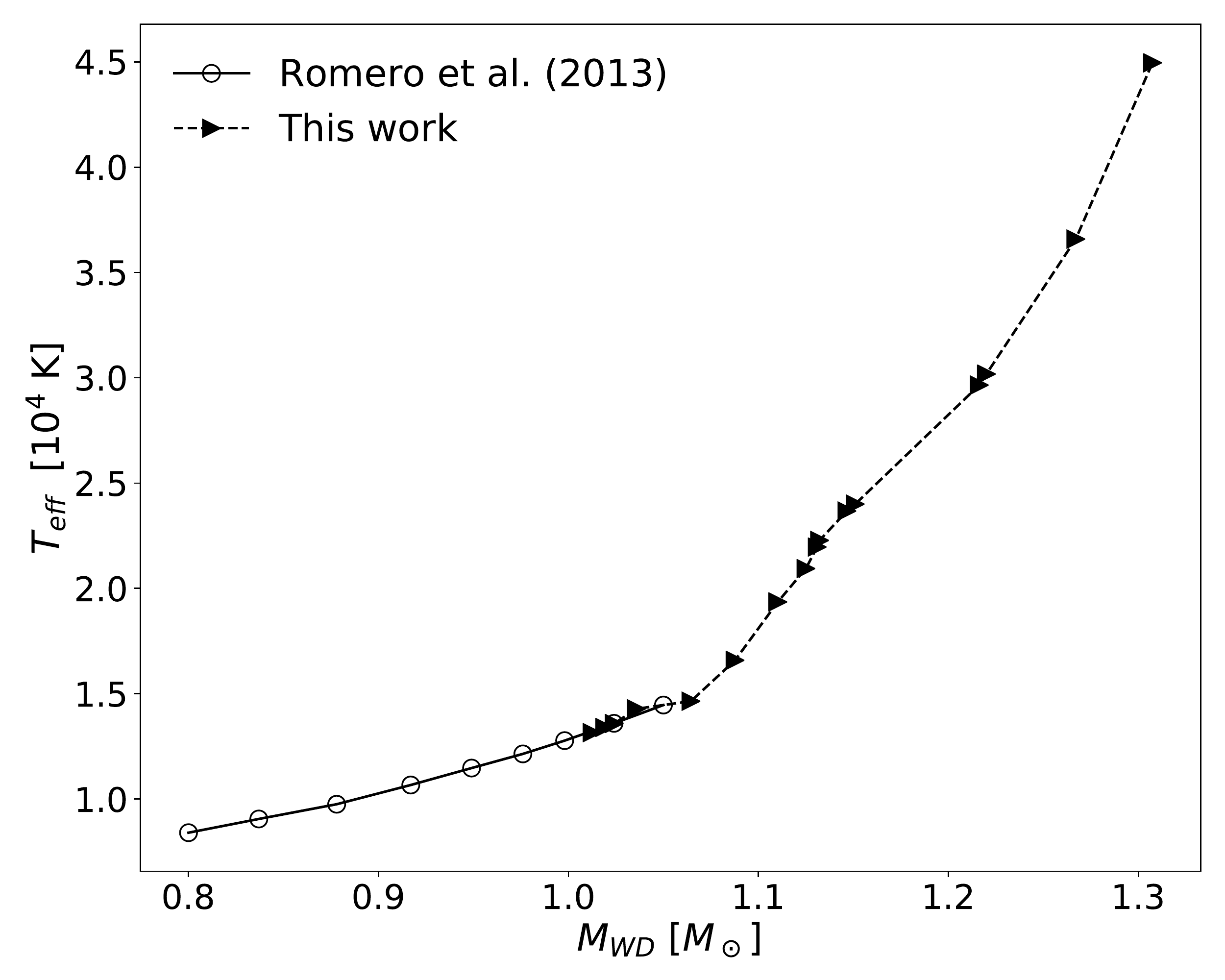}
    \caption{Comparison of effective temperature on onset of crystallization for a range of white dwarfs mass. Effective temperature on units of $10^4$ K. Circles are the data extracted from \citet{Romero2013} and triangles are results from this work. There is a agreement between both works for values of mass $M_{\rm WD} \sim 1$ M$_{\rm \odot}$ and the effective temperature raises as mass also increases.}
    \label{fig:da_teff_comparison}
\end{figure}

We compare the values of the effective temperature at the onset of crystallization, as a function of the white dwarf mass, with the results obtained by \citet{Romero2013} in Figure \ref{fig:da_teff_comparison}. Note that the results obtained in this work (triangles) are in good agreement with those from \citet{Romero2013} (circles), in the stellar mass range where both works overlap.

\begin{table}
	\centering
	\caption{Age and effective temperature on the onset of crystallization for different Coulomb coupling parameter. The age of crystallization is defined as time starting at the cooling sequence and ending at the onset of crystallization and is given in Gyr. Effective temperature in Kelvin. The effective temperature for $\Gamma_{\rm full} = 220$ on the last row agrees with table 2 of \citet{Romero2013}.}
	\label{tab:diff_in_age_for_gammas}
	\begin{tabular}{c|cc|cc}
	\hline
	& \multicolumn{2}{c}{$\Gamma_{\rm full} = 175$} & \multicolumn{2}{c}{$\Gamma_{\rm full} = 220$} \\
	$M_{\rm WD}$ [M$_{\rm \odot}$] & $T_{\rm eff}$ [K]& Age [Gyr] & $T_{\rm eff}$ [K]& Age [Gyr]\\
	\hline
	1.012 & 15095 & 0.647 & 13152 & 0.871 \\
	1.019 & 15376 & 0.628 & 13396 & 0.848 \\
	1.024 & 15552 & 0.678 & 13584 & 0.925 \\
	\hline
	\end{tabular}
\end{table}

\subsection{White dwarf cooling times}
\label{cooling}

The cooling time is defined as the time spent by a star in the white dwarf cooling sequence ($t_{\rm cool}$). We define the beginning of the cooling sequences as the point of maximum effective temperature in the post-AGB stage, before the star enters the cooling sequence. The progenitor age ($t_{\rm i}$) is defined as the time from the ZAMS to the beginning of the cooling sequence. Then the total age of a white dwarf star can be computed as $t_{\rm total}=t_i + t_{\rm cool}$. Table \ref{tab:ages_teff_cristal} summarizes the characteristic time-scales for all the sequences computed in this work. Specifically, for each sequence we listed the initial mass in the ZAMS (column 1) and the final mass at the cooling curve (column 2), along with the age for hydrogen and helium exhaustion at the core in columns 3 and 4, respectively. The progenitor age $t_i$ is listed in column 5. The age for hydrogen and helium depletion and the progenitor age are equal for both hydrogen and helium atmosphere white dwarfs sequences, since both have evolved equally on the stages prior to the cooling sequence. We also listed the effective temperature and the age in the cooling curve at the onset of crystallization for the hydrogen (columns 6 and 7) and helium atmosphere (columns 9 and 10) sequences. Finally, the cooling times corresponding to an effective temperature of $10\, 000$ K are listed in columns 8 and 12 for sequences with hydrogen and helium atmospheres, respectively.

\begin{table*}
\centering
 	\caption{Mass at ZAMS, white dwarf mass, age when hydrogen is depleted at center ($t_{\rm H}$), age when helium is totally consumed at the center ($t_{\rm He} $),  progenitor ages ($t_{\rm i}$), effective temperature on crystallization $(T_{\rm eff_c})$ and age at crystallization ($t_{\rm cryst}$, accounted from the beginning of the cooling sequence) and cooling time ($t_{\rm cool} = $ age at $T_{\rm eff} = 1\times10^4$ K minus $t_i$) for H and He atmosphere white dwarfs. Ages ($t_{\rm H}$, $t_{\rm He}$, $t_{\rm i}$) in units of Myr and crystallization and cooling times on units of Gyr.}
 	\label{tab:ages_teff_cristal}
\begin{tabular}{ccccc|ccc|ccc}
\hline
&  & & & & \multicolumn{3}{c}{H-atmosphere} & \multicolumn{3}{c}{He-atmosphere} \\
$M_{\rm ZAMS}$ &  $M_{\rm WD}$ & $t_{\rm H}$ & $t_{\rm He}$ & $t_i$ & $T_{\rm eff_c}$  &  $t_{\rm cryst}$ &$t_{\rm cool}$ & $T_{\rm eff_c}$  &  $t_{\rm cryst}$ &  $t_{\rm cool}$  \\

[M$_{\rm \odot}$] & [M$_{\rm \odot}$] & [Myr] & [Myr] & [Myr] & [K] & [Gyr] & [Gyr] & [K] & [Gyr] & [Gyr] \\
\hline
8.80    & 1.012 &  24.99  & 27.87  & 28.53  & 13152 &  0.870 &  3.65  &  13738  & 0.839 &    2.99 \\
8.90    & 1.019 &  24.50  & 27.30  & 27.94  & 13396  & 0.847 &  3.69  &  13837  & 0.839 &    2.99 \\
9.00    & 1.024 &  23.88  & 26.61  & 27.26  & 13584   &0.924 &  4.09  &  13634  & 0.933 &    3.48 \\
9.20    & 1.036 &  22.99  & 25.73  & 26.24  & 14281  & 0.810  &  3.75  &  14714  & 0.756 &    3.36 \\
9.50    & 1.064 &  21.63  & 24.17  & 24.63  & 14643&   0.754 &  3.13  &  14098  & 0.924 &    2.85 \\
9.80    & 1.088 &  20.49  & 22.82  & 23.24  & 16585 &  0.636 &  3.39  &  15858 &  0.742 &    2.90 \\
10.00   & 1.11  &  19.75  & 22.03  & 22.45  & 19356  & 0.415 &  3.04  &  17945  & 0.538 &    2.62 \\
10.10   & 1.125 &  19.42  & 21.64  & 22.03  & 20941   &0.338 &  2.65  &  19552   &0.405 &    2.34 \\
10.20   & 1.131 &  19.06  & 21.23  & 21.62  & 21963 &  0.310 &  2.72  &  19815&   0.387 &    2.48 \\
10.30   & 1.132 &  18.75  & 20.87  & 21.26  & 22276  & 0.296 &  2.71  &  20383 &  0.376 &    2.57 \\
10.45   & 1.147 &  18.35  & 20.37  & 20.74  & 23673   &0.253 &  2.52  &  23006  & 0.285 &    2.21 \\
10.50   & 1.151 &  18.19  & 20.17  & 20.54  & 24001&   0.250 &  2.51  &  23339&   0.278 &    2.21 \\
10.80   & 1.216 &  17.38  & 19.31  & 19.55  & 29651 &  0.183 &  2.28  &  29982 &  0.181 &    1.91 \\
11.00   & 1.22  &  16.83  & 18.63  & 18.88  & 30177  & 0.177 &  2.25  &  30244  & 0.179 &    1.90 \\
11.40   & 1.267 &  15.86  & 17.51  & 17.73  & 36584   &0.133 &  1.93  &  36808 &  0.132 &    1.71 \\
11.80   & 1.307 &  15.02  & 16.55  & 16.75  & 44953  & 0.098 &  1.54  &  45137  & 0.099 &    1.31 \\
\hline
\end{tabular}
\end{table*}

\begin{figure}
    \centering
    \includegraphics[width=\columnwidth]{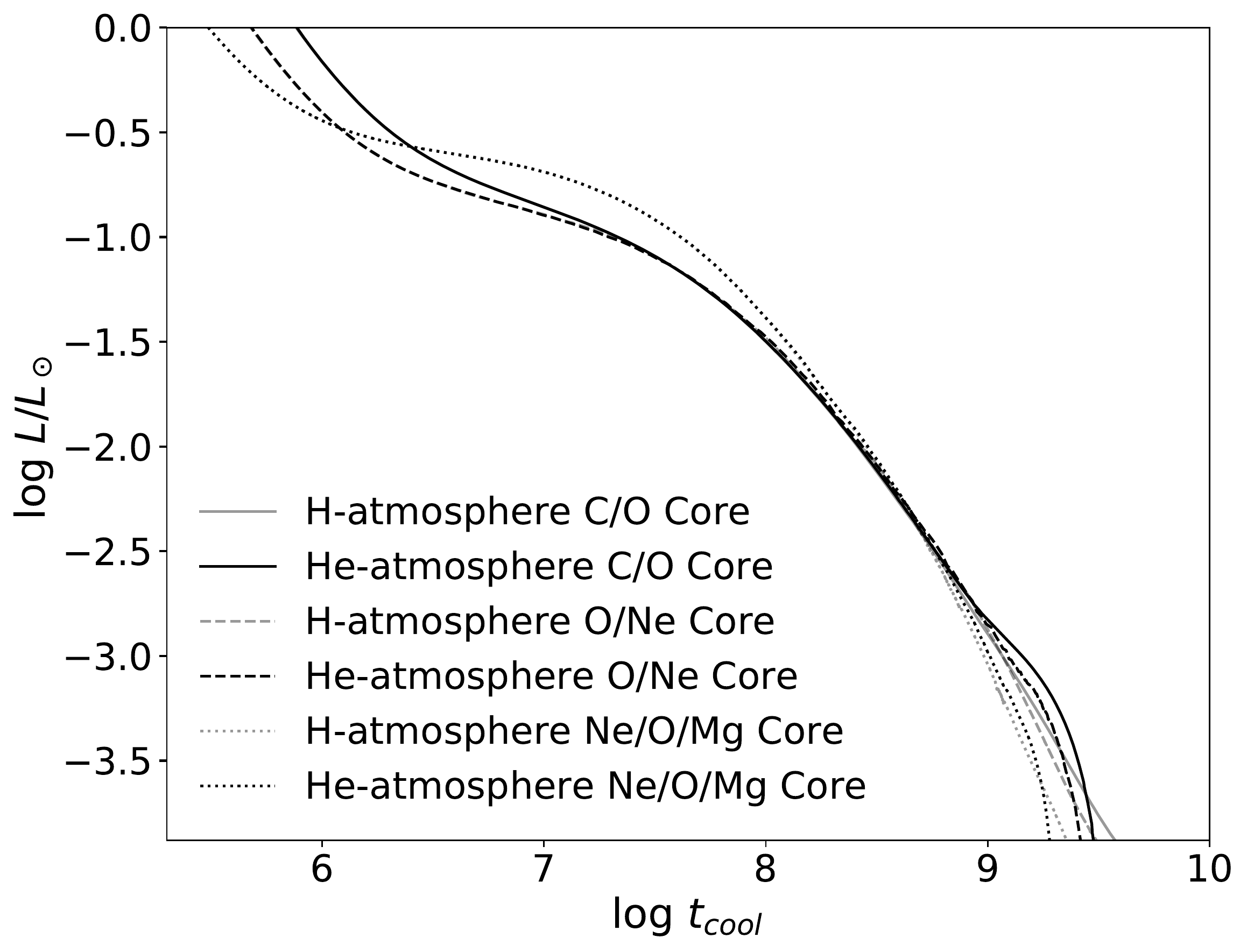}
    \caption{Cooling times for selected H and He atmosphere cooling sequences. We present 3 sets of H and He atmosphere sequences: a C/O core set with mass 1.019 M$_{\rm \odot}$ as solid line. A O/Ne set with mass 1.11 as dashed line and a Ne/O/Mg set with mass 1.22 M$_{\rm \odot}$ as dotted line. Gray lines represent H atmosphere and black lines are He atmosphere white dwarfs sequences. The final chemical profile for the H atmosphere are shown in Figure \ref{fig:comp_chemical_profile_da}.}
    \label{fig:tcool_log}
\end{figure}

\begin{figure}
    \centering
    \includegraphics[width=\columnwidth]{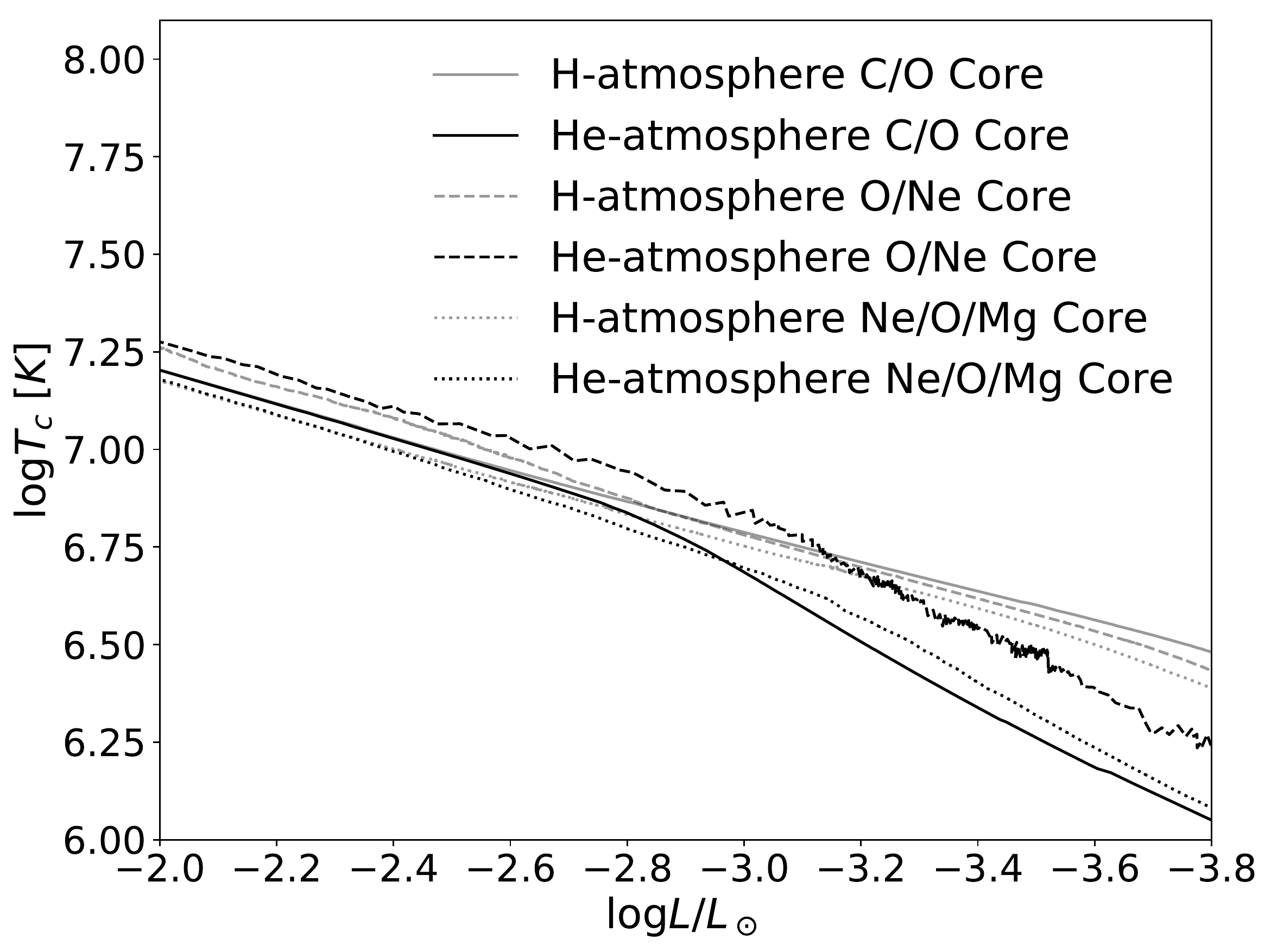}
    \caption{Logarithm of central temperature in terms of star luminosity. Three sets of H and He atmosphere are shown, the same as Figure \ref{fig:tcool_log}. Faint lines represent H atmosphere and strong lines are He atmosphere white dwarfs.}
    \label{fig:tc_lum}
\end{figure}

Figure \ref{fig:tcool_log} shows the cooling times for H and He atmosphere sequences with the same stellar mass, 1.019 (solid line), 1.11 (dashed line) and 1.22 $M_{\odot}$ (dotted line), whose chemical profiles are presented in Figure \ref{fig:comp_chemical_profile_da}. Gray lines represent hydrogen atmosphere and black lines correspond to helium atmosphere sequences.
For luminosities $\log L/L_{\rm \odot} \gtrsim -2.8$, He-atmosphere sequences cool somewhat slower then the H-atmosphere counterparts. On the other hand, for $\log L/L_{\rm \odot} \lesssim -2.8$ He-atmosphere sequences cools considerably faster leading to differences in cooling age of $\sim 1$ Gyr for luminosities of $\log L/L_{\rm \odot} \sim -4$.
These behavior can be explain by means of the central temperature of the star. Figure \ref{fig:tc_lum} depicts the central temperature $T_{\rm c}$, in a logarithmic scale, in terms of the surface luminosity for the sequences presented in Figure \ref{fig:tcool_log}.

On luminosities higher than $\log L/L_{\rm \odot} \approx -3$ the central temperature for the He atmosphere sequences is slightly larger than in the case of a H-atmosphere sequence with the same mass. However, for lower luminosities there is a large difference in $T_{\rm c}$ between H and He atmosphere sequences. The release of internal energy due to crystallization occurs at higher luminosities for He atmosphere sequences, delaying the cooling for those luminosities. Since helium is more transparent than hydrogen, He-atmosphere sequences lose energy faster than H atmosphere sequences at lower luminosities.
The results presented in Figure \ref{fig:tcool_log} agrees with the results from \citet{Camisassa2017} (see their Figures 7 and 11), who calculated evolutionary sequences for lower mass white dwarfs ($M_{\rm WD} \leq 1$ M$_{\rm \odot}$) with hydrogen-deficient atmosphere and compared against hydrogen rich white dwarfs from \citet{Camisassa2016}.

The core composition also impacts the cooling times of white dwarfs stars. \citet{Garcia-berro97-2} compared cooling times for sequences with same mass and two core compositions, C/O and O/Ne , and obtained a difference in cooling time of the order of $\sim 2$ Gyr. This is due to the difference in the heat capacity of the core, which is lower for a O/Ne mixture than for an C/O mixture, reducing the  cooling time for a core with a higher mean molecular weight.

The Debye regime of fast cooling for white dwarfs is important for very low luminosities. In particular, \citet{Althaus2007} found that the Debye cooling regime for H-atmosphere white dwarfs starts to be important for luminosities below $\log L/L_{\rm \odot} \sim -4$ for masses $M_{\rm WD} \geq 1.28$ M$_{\rm \odot}$. Since our computations end at those luminosities, the cooling times presented in Figure \ref{fig:tcool_log} are not affected by Debye cooling.

In the region of $0 > \log L/L_{\rm \odot} > -1.5$ the effects of neutrino losses are present and dominant, as can be see from Figure \ref{fig:tcool_log}. The neutrino emission decreases at higher luminosities for the more massive sequences when compared to lower masses sequences, however, there are no difference between the H and He atmospheres sequences at those luminosities. The rate of neutrino emission becomes negligible at $\log L/L_{\rm \odot} \sim -1$ for the sequences with masses $1.019$ M$_{\rm \odot}$ and $1.11$ M$_{\rm \odot}$, and at $\log L/L_{\rm \odot} \sim -0.5$ for the sequence with $1.22$ M$_{\rm \odot}$, in agreement with the results from  \citet{Althaus2007}. However, the evolution for luminosities lower than $\log L/L_{\rm \odot} \sim -2$ in our computations differs from the ones computed by \citet{Althaus2007}. In our computations we consider a crystallization treatment that mimics the results from the phase diagram presented by \citet{Horowitz2010}, while \citet{Althaus2007} considers the phase diagram from \citet{Segretain1994}, which increases the crystallization temperature by $\sim 2\, 000$ K (see \citet{Romero2013} for details). Also  \citet{Althaus2007} considers a fixed O/Ne core chemical profile for all stellar masses, while in the cooling sequences calculated in this work, the chemical profiles come from evolutionary computations and are consistent with the stellar mass and covers C/O, O/Ne and Ne/O/Mg cores.
The difference in the cooling times between our computations and those from \citet{Althaus2007}, are  $\sim 0.6$ Gyr at $\log L/L_{\rm\odot} \sim -4$ for a stellar mass of $\sim 1.21 M_{\rm \odot}$.

\section{Mass-Radius Relation}
\label{sec:mass_radius}

\begin{figure}
	\centering
	\includegraphics[width=\columnwidth]{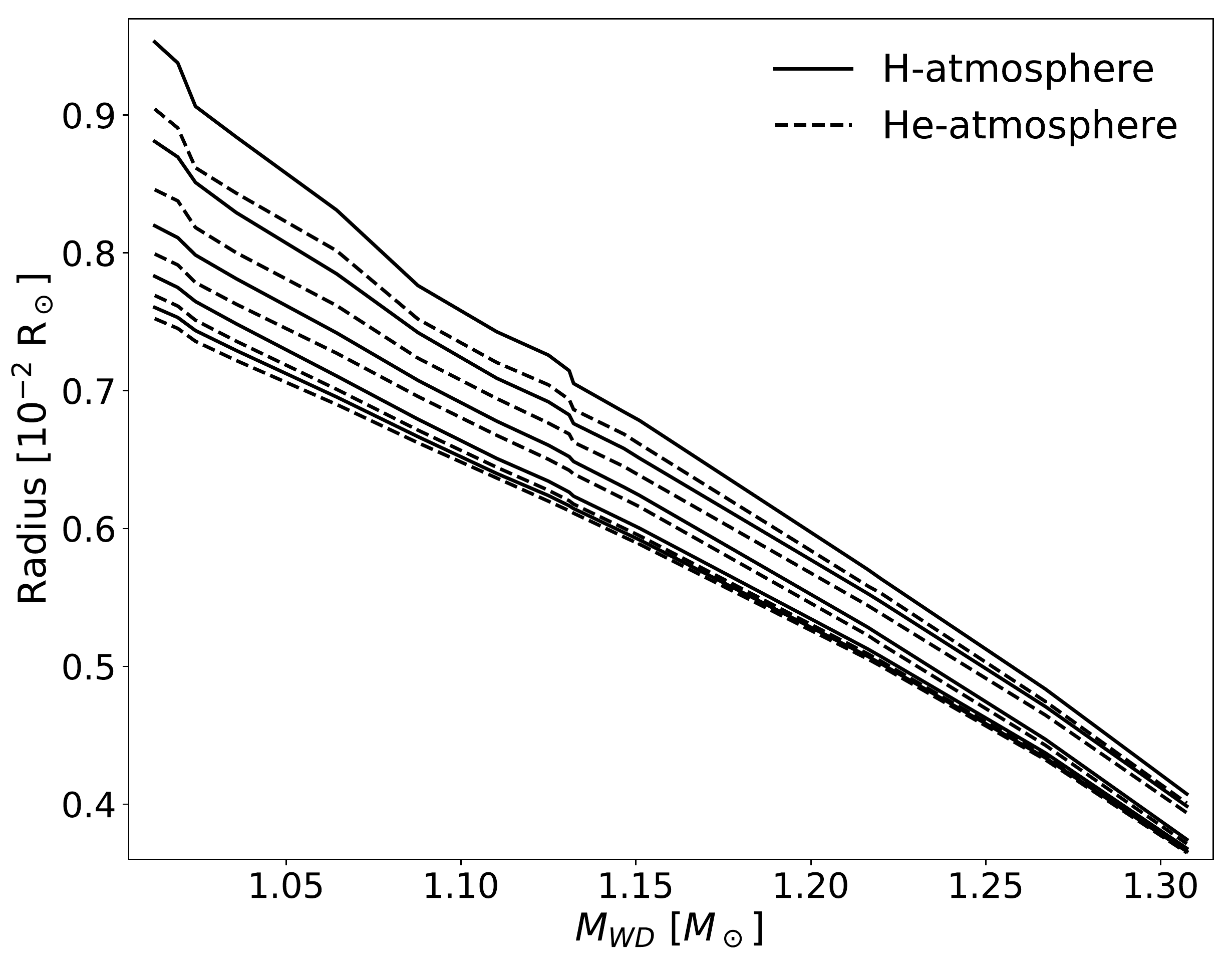}
	\caption{Mass-Radius relation for H and He atmosphere white dwarfs models, represented as solid and dashed lines, respectively. From top to bottom, the lines represent effective temperatures of $150\, 000$ K, $100\, 000$ K, $50\, 000$ K, $25\, 000$ K and $10\, 000$ K. See text for details.}
	\label{fig:mass-radius}
\end{figure}

Figure \ref{fig:mass-radius} shows the mass-radius relation of our sequences with H and He atmosphere, as solid and dashed lines, respectively. We consider effective temperatures of $150\, 000$ K, $100\, 000$ K, $50\, 000$ K, $25\, 000$ K and $10\, 000$ K, shown from top to bottom.
The rate of contraction is higher for high effective temperatures compared to low effective temperatures. Due to their higher surface gravity, more massive sequence show a smaller range in the value of the radius for the different effective temperatures considered here.
For a sequence with H atmosphere characterized by a stellar mass of $1.267$ M$_{\rm \odot}$ the total radius changes from $4.83\times10^{-3}$ R$_{\rm \odot}$ to $4.34\times10^{-3}$ R$_{\rm \odot}$ when the star cools from $150\, 000$ K to $10\, 000$ K. On the other hand, for a sequence with a  mass of $1.064$ M$_{\rm \odot}$, the radius decreases from $8.31\times10^{-3}$ R$_{\rm \odot}$ to $6.95\times10^{-3}$ R$_{\rm \odot}$ for the same range of effective temperatures. This means a change in radius of $\sim 10\%$ for the more massive and $\sim 16\%$ for the latter. Those values change when sequences with He atmosphere are considered. For a He atmosphere sequence with $1.267$ M$_{\rm \odot}$ the radius change from $4.74\times10^{-3}$ R$_{\rm \odot}$ to $4.32\times10^{-3}$ R$_{\rm \odot}$ ($\sim 9\%$) and for the $1.064$ M$_{\rm \odot}$ sequence there is a change from $8.02\times10^{-3}$ R$_{\rm \odot}$ to $6.90\times10^{-3}$ ($\sim 14\%$) when considering the same effective temperature range. Note that, although the contraction of H atmosphere sequences are slightly higher, their final radii are higher compared to the He atmosphere counterparts, showing that a small amount of hydrogen of the order of $M_{\rm H} \approx 10^{-6}$ M$_{\rm WD}$ can change the radius by $5.4\times10^{-5}$ R$_{\rm \odot}$ ($\sim 1 \%$) for the $M_{\rm WD} = 1.064$ M$_{\rm \odot}$ sequence.

\begin{figure}
	\centering
	\includegraphics[width=\columnwidth]{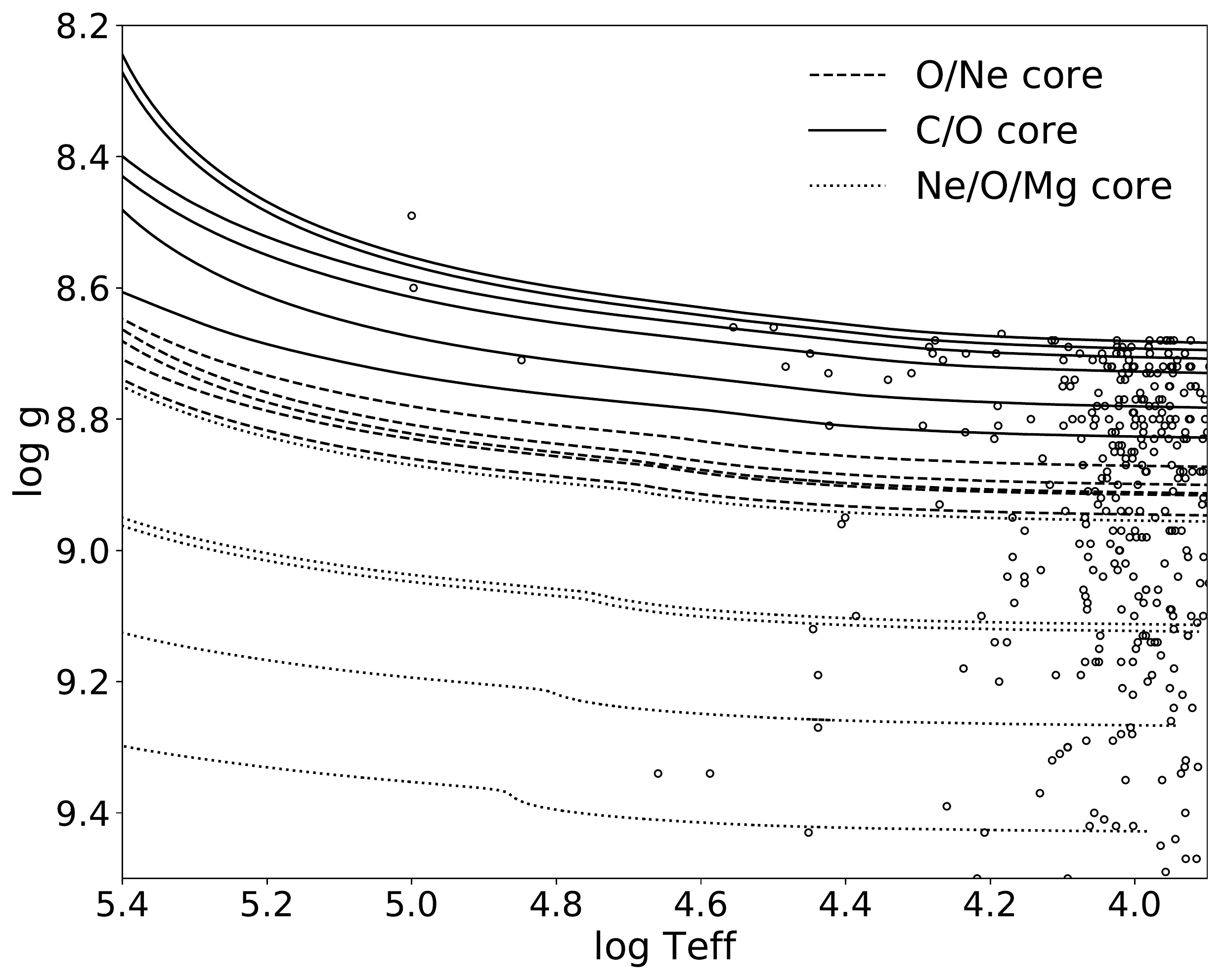}
	\caption{$\log g$ - $T_{\rm eff}$ relation for H atmosphere white dwarfs. $T_{\rm eff}$ from $\approx 250\, 000$ K to $\approx 10\, 000$ K. Empty circles correspond to H atmosphere white dwarf stars from the SDSS catalog presented in \citep{Kepler2016}. See text for details.}
	\label{fig:da_logg-tef}
\end{figure}

\begin{figure}
	\centering
	\includegraphics[width=\columnwidth]{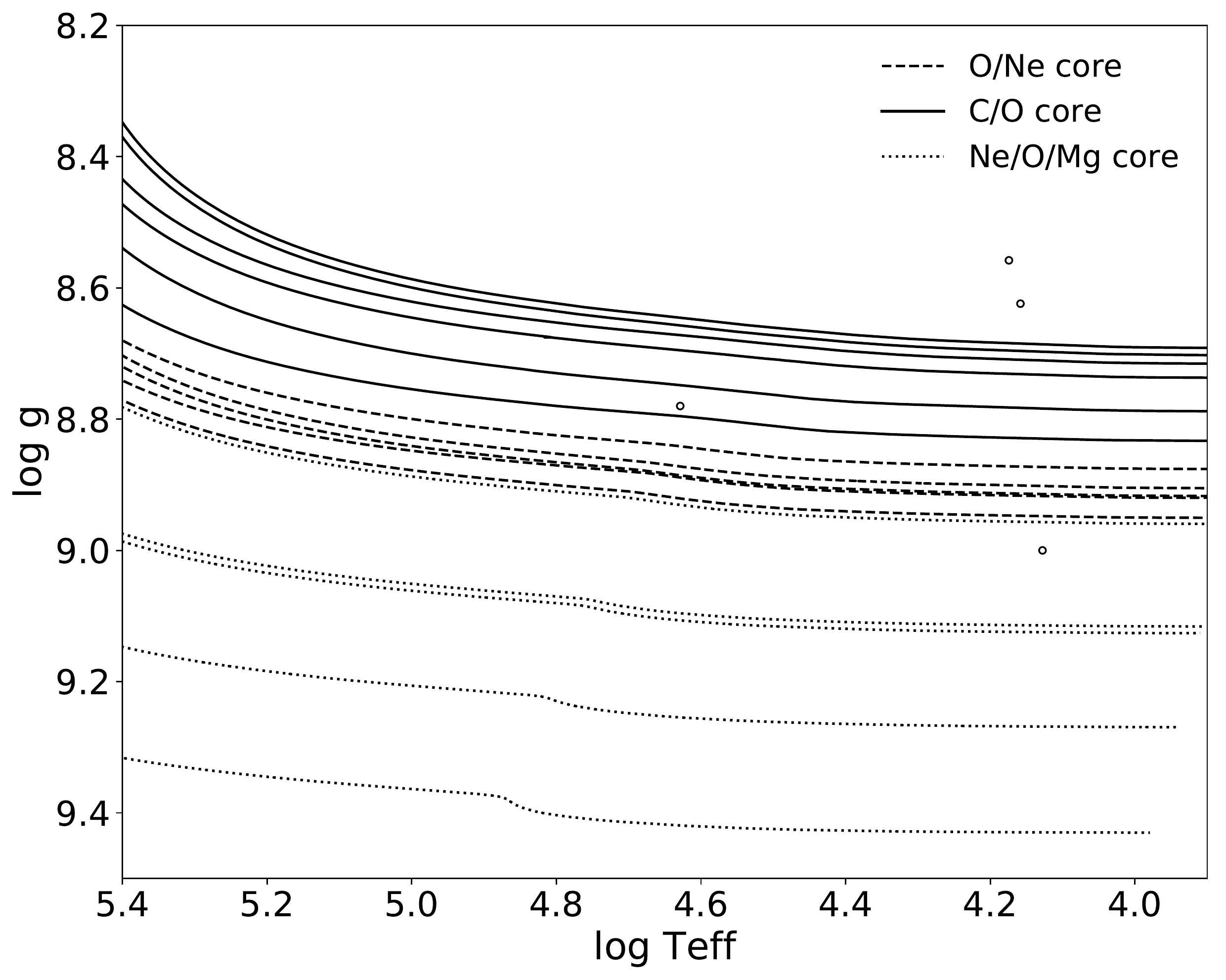}
	\caption{$\log g$ - $T_{\rm eff}$ relation for He atmosphere white dwarfs. $T_{\rm eff}$ from $\approx 250\, 000$ K to $\approx 10\, 000$ K. Empty circles correspond to He atmosphere white dwarf stars from the SDSS catalog presented in \citep{Kepler2016}. See text for details.}
	\label{fig:db_logg-tef}
\end{figure}

The surface gravity $\log g$ as a function of effective temperature $T_{\rm eff}$ for H and He atmosphere sequences are presented in Figures \ref{fig:da_logg-tef} and \ref{fig:db_logg-tef}, respectively.
The solid lines correspond to C/O cores sequences while dashed and dotted lines correspond to O/Ne and Ne/O/Mg core sequences, respectively. Stellar mass decreases from top to bottom. Also plotted, as empty circles, are a selection of white dwarfs with spectroscopic masses $M_{\rm WD} > 1$ M$_{\rm \odot}$ from the SDSS DR12 catalog presented by \citep{Kepler2016}.
Surface gravities and effective temperatures for H-atmosphere white dwarfs are corrected to 3D convection \citep{Tremblay2013} and have mean values of $\left< \log g \right> = 8.987 \pm 0.21$ and $\left< T_{\rm eff} \right> = 11\, 591 \pm 501$. From Figure \ref{fig:da_logg-tef} we find a good agreement between observations and our theoretical cooling sequences with masses $M_{\rm WD} \leq 1.151 $ M$_{\rm \odot}$.
Comparing Figures \ref{fig:da_logg-tef} and \ref{fig:db_logg-tef} we can see that the sequences are slightly shifted to smaller gravities, on average $\Delta \log g = 0.0135$, for lower effective temperatures due to contraction, which reduced the radius by $\Delta R = 3.246\times10^{-5}$ $R_{\rm \odot}$ on average. The increase in $\log g$ at $\log T_{\rm eff} \sim 4.8 - 4.9$ presented in figures \ref{fig:da_logg-tef} and \ref{fig:db_logg-tef} for the Ne/O/Mg sequences with $M_{\rm WD} > 1.2$ M$_{\rm \odot}$ is due to the decrease in the neutrino luminosity, which causes a reduction in the radius of the model. This behavior agrees with the results from \citet{Althaus2005} who were the first to report this phenomenon.

\begin{figure}
	\centering
	\includegraphics[width=\columnwidth]{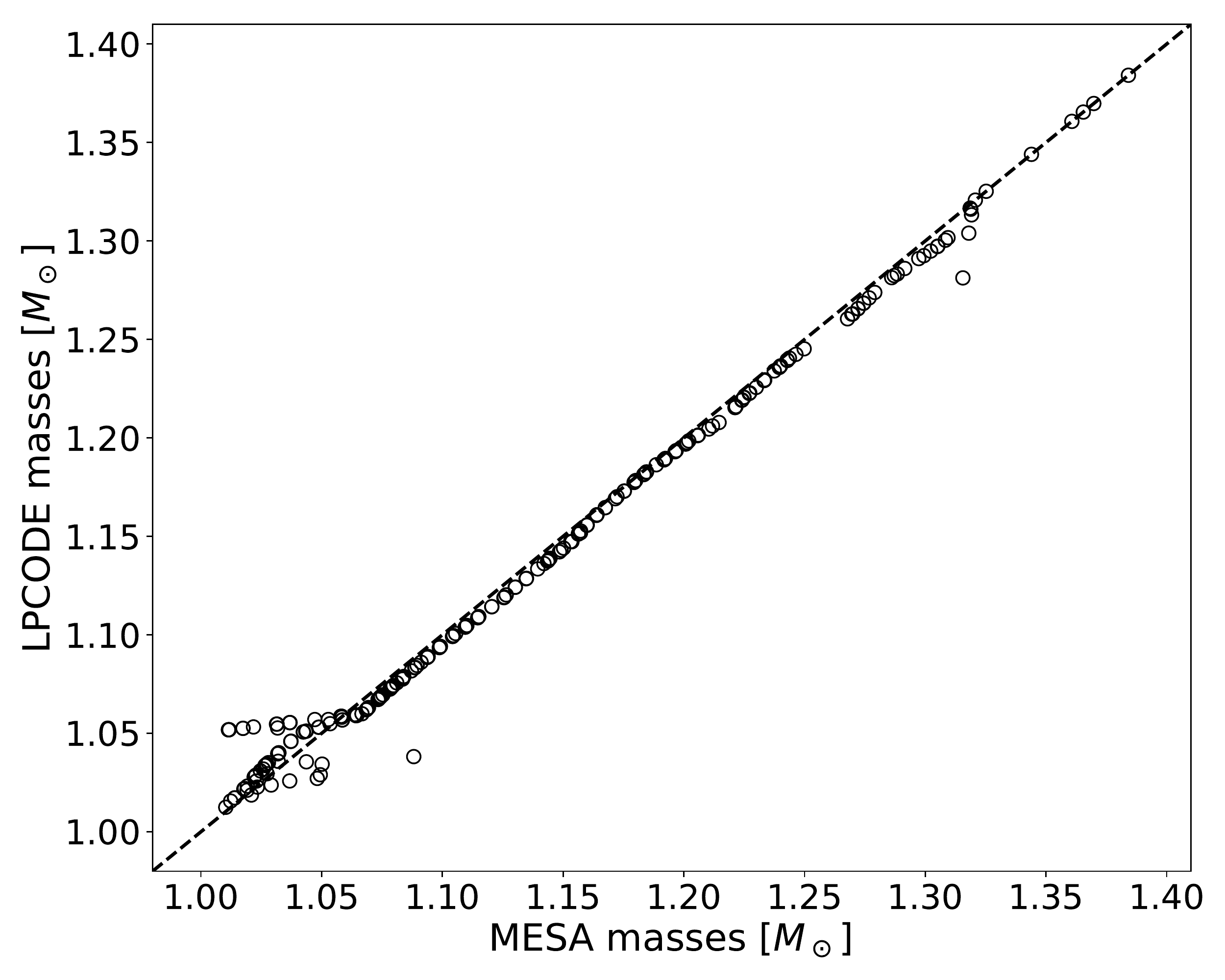}
	\caption{Comparison between the stellar mass for a sample of 252 spectroscopically identified DA white dwarf stars from the SDSS DR12 catalog, computed using the MESA sequences (x-axis) and LPCODE sequences (y-axis). The dashed line indicates the 1:1 correspondence. }
	\label{fig:masses_comp}
\end{figure}

For comparison proposes, we compute the stellar mass of a sample of 252 stars from the SDSS DR12 catalog of \citet{Kepler2016} using the cooling sequence computed in this work and those obtained using the LPCODE evolutionary code \citep{Althaus2005-lpcode,Romero2013}. The results are shown in Figure \ref{fig:masses_comp} where we depict the stellar mass obtained from the MESA tracks against the mass computed using the LPCODE tracks. As can be seen, there is practically no difference between the two determinations. The deviations from the 1:1 correspondence for stellar masses around $1.05 M_{\rm \odot}$ are due to the transition from a C/O to a O/Ne core composition, which occurs at $1.05 M_{\rm \odot}$ for the LPCODE, while in our simulations it occurs at $1.088 M_{\rm \odot}$.

\section{Conclusion}
\label{sec:7-conclusion}

In the present work we have calculated full evolutionary sequences with $Z=0.02$ for massive H and He atmosphere white dwarf stars using the evolutionary code MESA version r8845. Our sequences started from the zero-age main sequence with masses from 8.8 to 11.8 M$_{\rm \odot}$, evolving through central hydrogen and helium burning, thermally pulsing and mass-loss phases ending as white dwarfs on the cooling sequence. The resulting white dwarfs sequences show stellar masses in a  range of $M_{\rm WD} = 1.012 - 1.307$ M$_{\rm \odot}$.
We presented full chemical profiles, core mass range, cooling times with crystallization effects, mass-radius relation and initial-to-final mass relation. We also calculated masses from our evolutionary tracks for a selection of 252 massive stars from the SDSS DR12 catalog \citep{Kepler2016} and compared with derived masses from other evolutionary tracks presented on the literature.

Unlike models that only evolve to the AGB, our white dwarf sequences considers the entire evolutionary history of the corresponding progenitors, leading to realistic chemical profiles which are also consistent with stellar mass.
Note that the amount of hydrogen and helium affects both the age and mass determinations of DA and non-DA white dwarfs. The final mass range of our sequences cover the entire range of C/O, O/Ne and Ne/O/Mg core composition and also accounts for hybrids C/O-O/Ne white dwarfs. Reliable chemical composition of core and the chemical transition  regions presented on Figure \ref{fig:comp_chemical_profile_da} are of relevance for asteroseimological studies of pulsating white dwarf. The set of tracks presented here is the first in the literature to cover those core composition for massive white dwarfs, considering the evolutionary history of the progenitors, and presenting the entire chemical profiles for the sequences.

Our main results are the following:

\begin{enumerate}
\item From our simulations we obtained white dwarf sequences characterized with C/O, O/Ne and Ne/O/Mg cores being: C/O cores for $M_{\rm WD} \leq 1.088 $ M$_{\rm \odot}$, O/Ne cores for $1.088 < M_{\rm WD} \leq 1.147 $ M$_{\rm \odot}$, Ne/O/Mg cores for $M_{\rm WD} > 1.147 $ M$_{\rm \odot}$. Note that these values correspond to sequences with initial metallicity $Z=0.02$. We expect them to vary for different initial metallicities \citep{Doherty2015} . We compute a sequence with $M_{\rm WD} = 1.13$ M$_{\rm \odot}$ and $Z=0.015$ and found that the chemical structure does not change considerably from the sequence with $Z=0.02$ and a similar stellar mass in the cooling curve.

\item All sequences with $M_{\rm WD} \geq 1.024$ M$_{\rm \odot}$ experienced inward propagating carbon flames with subsequent helium burning of the outer shells, in the post-AGB stage. The carbon flame for sequences with masses $ 1.024 \leq M_{\rm WD} \leq 1.147$ M$_{\rm \odot}$ does not reach the central regions, forming a hybrid C/O-O/Ne white dwarf. For sequences with  $M_{\rm WD} > 1.147$ M$_{\rm \odot}$ the carbon flame reaches the center of the model producing a Ne/O/Mg core.
Our Hybrid models presents a O/Ne core surrounded by a triple-layer of oxygen-carbon-helium, differing from the work of \citet{Denissenkov2013} where they reported a C/O core surrounded by a O/Ne zone for hybrid C/O/Ne Super AGB stars.

\item Our choice of the Coulomb coupling parameter for crystallization $\Gamma_{\rm full} = 220$, based on the latest asteroseismology results, instead of the MESA default $\Gamma = 175$, can increase the cooling times by $\approx 0.26$ Gyr and decrease the effective temperature at the onset of crystallization by $\approx 2000K$.

\item The amount of hydrogen left at the envelope of a white dwarf has an important impact in determining the stellar mass and the cooling times. Even low quantities of hydrogen, as the ones expected for massive white dwarfs, can impact the radii and cooling times. In particular, our H-atmosphere sequences shows cooling times $\sim 0.23 - 0.66$ Gyr larger at $T_{\rm eff} \approx 10\, 000$ K, depending on stellar mass, when compared to He atmosphere sequence with the same mass.

\item The derived masses for a selection of 252 massive stars from the SDSS DR12 catalog \citep{Kepler2016} using our simulations are in fine agreement with masses computed using the evolutionary tracks of LPCODE \citet{Althaus2005-lpcode,Romero2012,Romero2013}, except around masses $M_{\rm WD} =1.05 - 1.088$ M$_{\rm \odot}$.
\end{enumerate}

\section*{Acknowledgements}

We acknowledge the valuable report of the anonymous referee. GRL acknowledges the support by CAPES-Brazil. ADR and SOK thank support from CNPq and PRONEX-FAPERGS/CNPq (Brazil). This research has made use of NASA's Astrophysics Data System.




\bibliographystyle{mnras}
\bibliography{ref.bib}



\appendix

\section{Chemical Profiles at the end of cooling sequence}
\label{sec:appendix-1}

In this section is presented the chemical profiles of all our models for H and He atmosphere white dwarfs. Figure \ref{fig:grid_1} shows the chemical profiles for H atmosphere on the column on the left and He atmosphere profiles on column on the right. Each row presents a H atmosphere and a He atmosphere models for a same mass. All chemical profiles presented in this section are for $T_{\rm eff} \approx 10,000$ K.

\begin{figure*}
    \centering
    \includegraphics[width=\linewidth, height=\textheight]{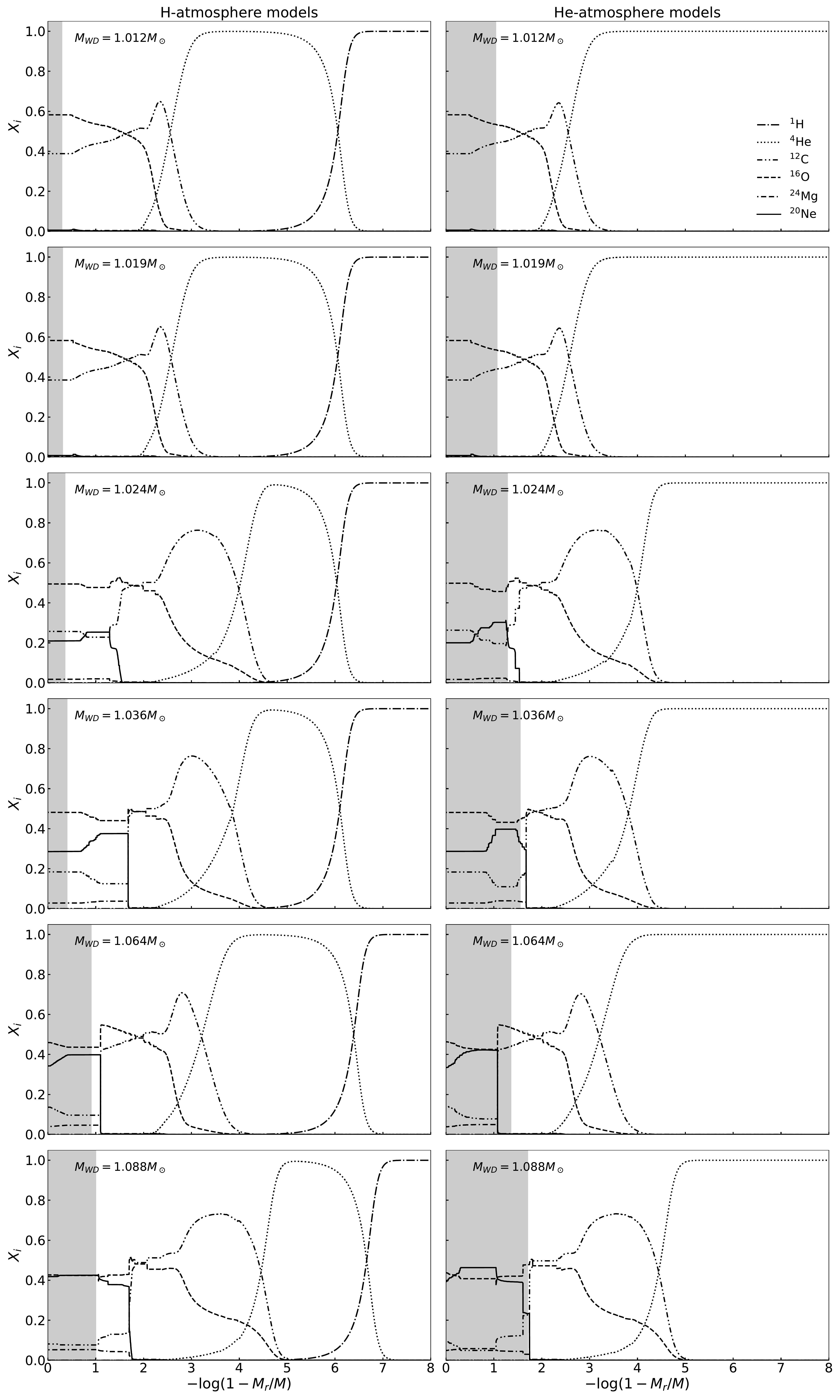}
    \caption{Chemical profiles in terms of outer mass fractions for H atmosphere (left) and He atmosphere (right) models at $T_{\rm eff} \approx 10\, 000$ K in the cooling sequence. The shaded regions means crystallized portion of core. Each row presents a H atmosphere (left) and a He atmosphere (right) for a same mass}
    \label{fig:grid_1}
\end{figure*}

\begin{figure*}
    \centering
    \includegraphics[width=\linewidth, height=\textheight]{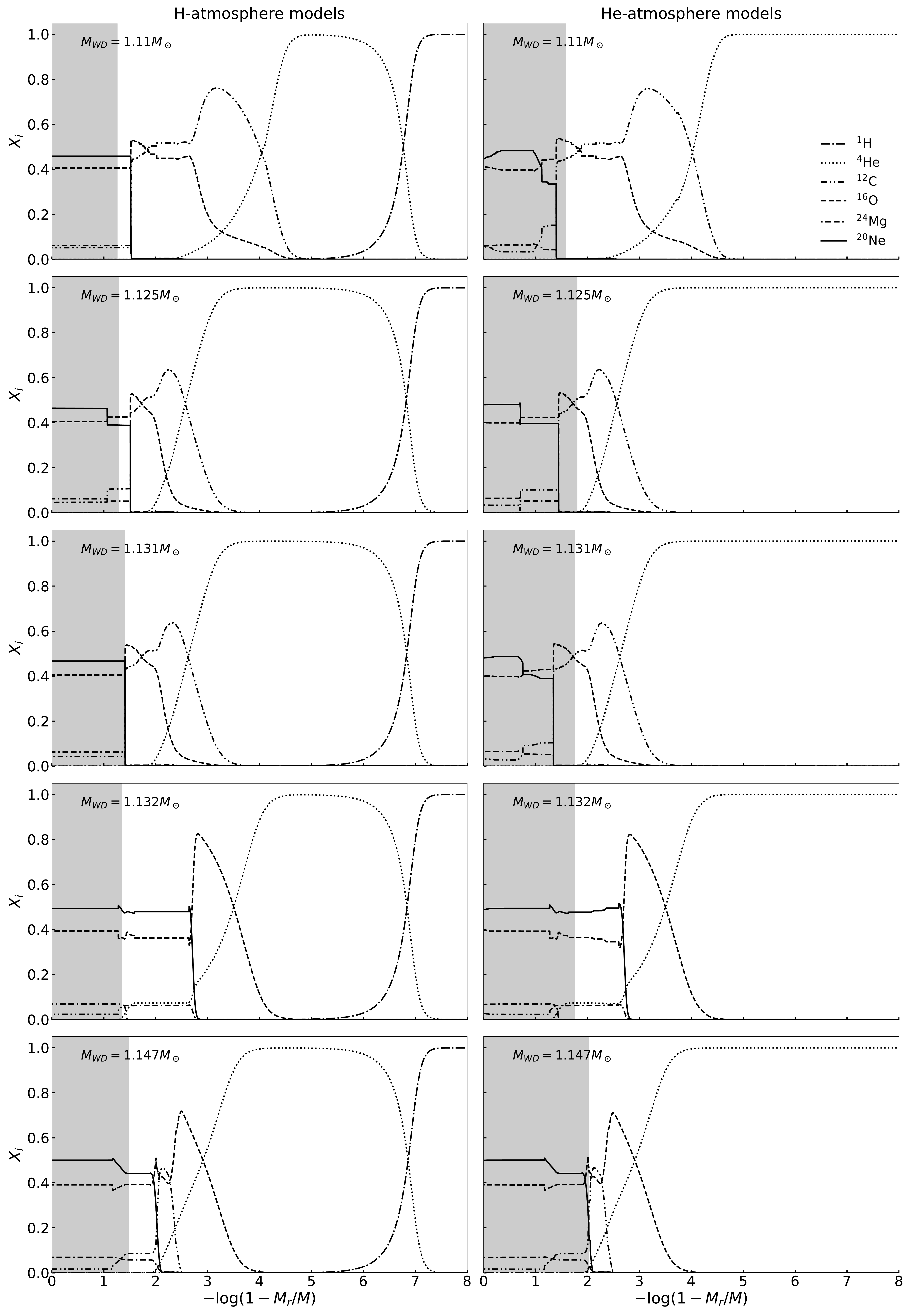}
    \contcaption{}
    \label{fig:grid_2}
\end{figure*}

\begin{figure*}
    \centering
    \includegraphics[width=\linewidth, height=\textheight]{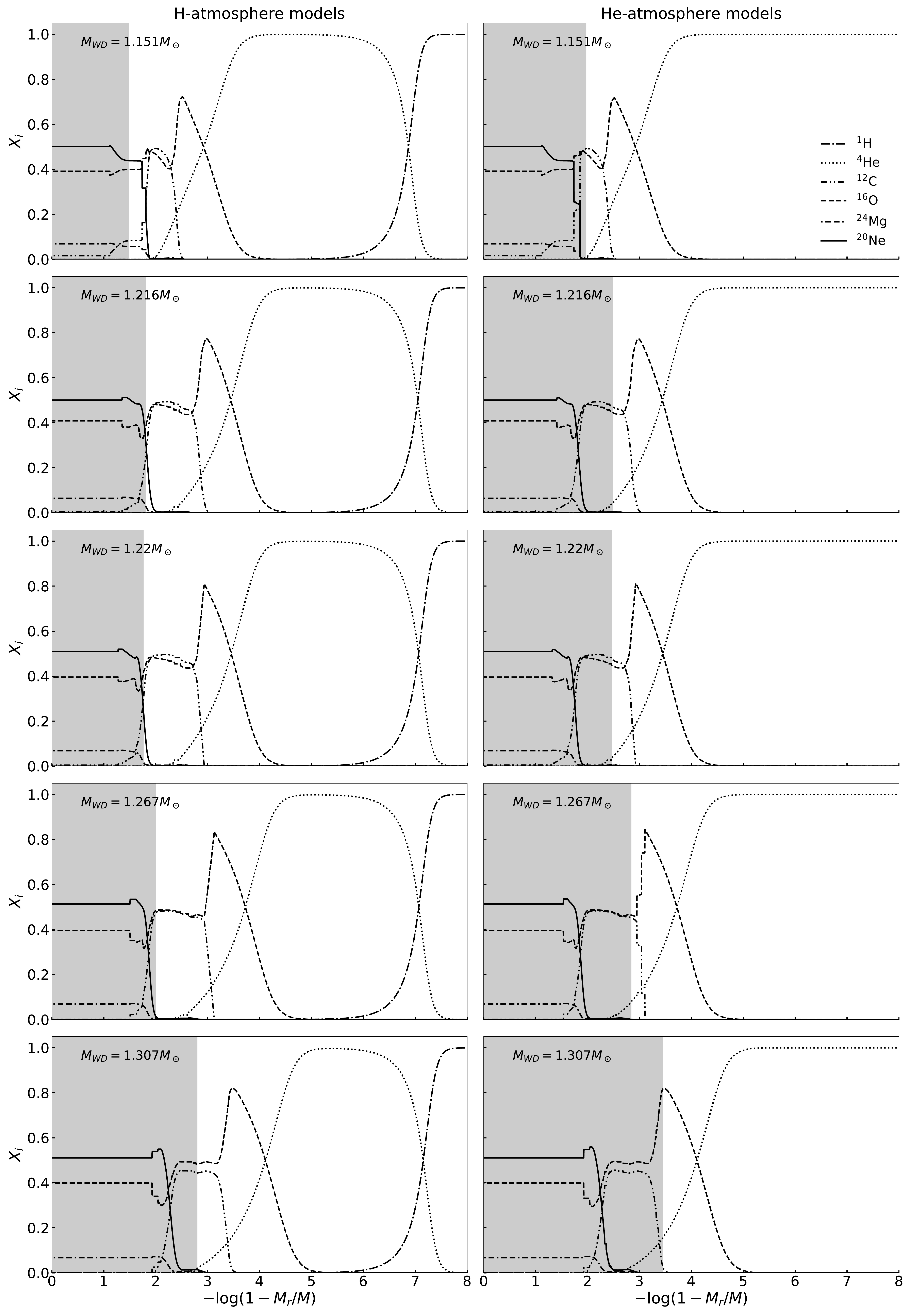}
    \contcaption{}
    \label{fig:grid_3}
\end{figure*}

\bsp	
\label{lastpage}
\end{document}